\documentclass[twocolumn,aps,prb,reprint,superscriptaddress,longbibliography]{revtex4-2}
\usepackage{graphicx}
\usepackage{multirow}
\usepackage{xr-hyper}
\usepackage{hyperref}
\usepackage{dsfont} 
\usepackage{pifont}
\hypersetup{
     colorlinks   = true,
     citecolor    = prlblue,
     linkcolor    = prlblue,
     urlcolor     = prlblue, 
}
\usepackage{physics}

\usepackage{blkarray}
\usepackage[dvipsnames]{xcolor}
\usepackage[utf8]{inputenc}
\usepackage[vietnamese, english]{babel}
\usepackage{chngcntr}
\usepackage{xcolor}
\usepackage{braket}
\usepackage{sidecap}
\usepackage{lipsum}
\usepackage{amssymb}
\usepackage{bbold}
\usepackage{booktabs}
\usepackage{geometry}
\usepackage{tikz}
\usetikzlibrary{arrows.meta}
\usepackage{xcolor}

\newcommand{\Nb}{N_b}

\begin{document}
\definecolor{prlblue}{HTML}{2e3092}

\title{Intrinsic spin-orbit coupling in graphene stacks.}

\author{Guillermo Parra-Martinez}
\affiliation{IMDEA Nanoscience, C/ Faraday 9, 28049 Madrid, Spain}
\author{Alejandro Jimeno-Pozo}
\altaffiliation[Present address: ]{Department of Physics and Astronomy, Uppsala University, Box 524, 751 20 Uppsala, Sweden}
\affiliation{Donostia International Physics Center, Paseo Manuel de Lardiz\'{a}bal 4, 20018 San Sebasti\'an, Spain}
\author{Ady Stern}
\affiliation{Department of Condensed Matter Physics, Weizmann Institute of Science, Rehovot 7610001, Israel}
\author{Jose Angel Silva-Guill\'en}
\email{joseangel.silva@imdea.org}
\affiliation{IMDEA Nanoscience, C/ Faraday 9, 28049 Madrid, Spain}
\author{Francisco Guinea}
\email{paco.guinea@imdea.org}
\affiliation{IMDEA Nanoscience, C/ Faraday 9, 28049 Madrid, Spain}
\affiliation{Donostia International Physics Center, Paseo Manuel de Lardiz\'{a}bal 4, 20018 San Sebasti\'an, Spain}

\begin{abstract} 
The effect of spin-orbit coupling (SOC) in monolayer graphene has been thoroughly studied in the past. 
Its impact is commonly disregarded since the electronic structure of graphene is almost unaffected.
Consequently, SOC has also been ignored in other graphitic systems.
In this work, we calculate the intrinsic SOC in stacks of graphene layers from a model that includes all the valence orbitals of carbon as well as the intra-atomic SOC.
The leading spin-orbit term in a graphene multilayer combines a Rashba-like spin-flip process with interlayer hopping.
We find that this intrinsic term modifies the phase diagram and Fermi surfaces of finite rhombohedral stacks, and it leads to a distinct density-dependence of the magnetization in weak magnetic fields.
This Letter shows that, unlike in the single-layer case, SOC has important consequences in multilayer graphitic systems and it should not be neglected when studying their physical properties.
\end{abstract}

\maketitle

\emph{Introduction--}
Spin-orbit coupling (SOC) in graphene and other graphitic compounds has been thoroughly studied due to its fundamental~\cite{kane_z_2005,avsar2020colloquium} and possible technological~\cite{Han2014,PesinMacDonald2012,RocheValenzuela2014,Roche2015,Zhang2024} consequences. 
The fact that carbon is a light element implies that the  intra-atomic, intrinsic, SOC in carbon compounds is smaller than in materials which contain heavier elements. 
In a graphene single-layer, symmetry constraints reduce further the effect of the intra-atomic coupling, as the leading spin order term projected onto the Dirac cones --which describe the low-energy electronic structure-- is proportional to the square of the intra-atomic coupling divided by an energy scale of the order of the separation between the $\sigma$ and $\pi$ bands of graphene~\cite{kane_quantum_2005}.
Since this separation is of the order of a couple of eV, the accepted value for SOC in single-layer graphene is in the $\mu$eV range~\cite{min_intrinsic_2006,huertas-hernando_spin-orbit_2006,yao_spin-orbit_2007,shichau-2019-resonance}.

SOC in graphite and multilayer graphene stacks has been theoretically investigated~\cite{dresselhaus_spin-orbit_1965,guinea_spinorbit_2010,liu_giant_2010,mccann_spin-orbit_2010,konschuh_tight-binding_2010,konschuh_theory_2012,kormanyos_intrinsic_2013}. 
Importantly, most estimates suggest that SOC in graphene multilayers is significantly increased with respect to its value in a single-layer. 
Indeed, recent experimental results in multilayer graphene stacks have shown that the SOC in rhombohedral stacks may be of the order $0.1 -1$ meV~\cite{arp_intervalley_2024,auerbach_isospin_2025}, much higher than the accepted value for single-layer graphene mentioned before. 

The renewed interest in rhombohedral multilayer graphene (RMG) has revealed complex phase diagrams for different number of layers, which combine in novel ways seemingly incompatible features, such as spin and valley polarization with superconductivity. These phases typically appear at temperatures below 1K, and due to the relatively defect-free nature of the samples they have been reproduced in a number of works~\cite{lee_multicomponent_2016,shi_electronic_2020,zhou_half_2021,zhou_superconductivity_2021,hagymasi_observation_2022,arp_intervalley_2024,han_large_2024,liu_spontaneous_2024,han_correlated_2024,winterer_ferroelectric_2024,zheng_switchable_2025,yang_impact_2025,xie_tunable_2025,kumar_superconductivity_2025,morissette_striped_2025,kumar_superconductivity_2025,holleis_fluctuating_2025,deng_superconductivity_2025,yang_magnetic_2025,han_signatures_2025,morissette_evidence_2025,li_fractional_2025,deng_superconductivity_2025,nguyen_hierarchy_2025,seo_family_2025,han_evidence_2026}.
Up to now, theoretical and experimental works have focused on the effect of an induced SOC: Placing a graphitic stack on a substrate with a strong SOC (e.g. a TMDC), induces an Ising or Rashba SOC that impacts the phase diagram and superconducting properties of the system~\cite{arora2020superconductivity,jimeno2023superconductivity,koh2024symmetry,zhumagulov2024emergent,zhumagulov2024swapping,parra-martinez_nematic_2025,raines2026superconductivity}. 

In this Letter we study the consequences of the intrinsic SOC on the electronic properties of finite rhombohedral stacks, with emphasis on its effect on the flat bands expected at the surfaces of these systems. 
We first analyze, using symmetry arguments, the different terms that arise in the tight-binding (TB) hamiltonian when adding intrinsic SOC.
By estimating their values, we show that the most important is the one that involves an intra and inter-atomic spin-orbit transition.
We then include this leading intrinsic SOC term into a continuum model and examine its impact on the complex phase diagram using Hartree-Fock (HF) of a particular system, four-layer rhombohedral graphene (R4G), a system that features exotic superconducting states~\cite{han_signatures_2025} and that we have studied recently~\cite{parra-martinez_band_2025,parra-martinez_nematic_2025}.
We find that in multilayer systems this term has an important impact on the stabilization of the different correlated phases.
We finally propose an experimental setup that would allow to measure the strength of the leading SOC term experimentally.
Our results show that, contrary to the single-layer case where it is neglected, intrinsic SOC should be included when studying graphitic stacks such as Bernal and Rhombohedral-stacked multilayer graphene, twisted bilayer graphene, and related twisted assemblies.

\begin{figure}
    \includegraphics[scale=1]{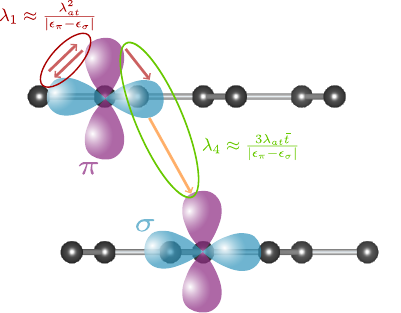}
    \caption{Sketch of second order perturbation processes which lead to $\lambda_1$ (red ellipse, Kane-Mele term), and $\lambda_4$ (green ellipse, spin-flip + layer hopping) in eq.~(\ref{hamilso_1}).}
    \label{fig:SOC_sketch}
\end{figure}

\emph{Tight-binding model for intrinsic spin-orbit coupling in graphene stacks--}
The spin-independent part of the TB hamiltonian of multilayer rhombohedral graphene and its parameters are well-known (see section S1 of the Supplemental Material (SM)~\cite{SM} for a detailed description).
To estimate the effective spin-orbit coupling we include the atomic SOC in our TB hamiltonian which includes the $2s$ and $2p$ orbitals of the individual carbon atoms. 
Conventionally, the bands in graphite are divided into $2 \times 2$ (spin) $\pi$ bands, associated to the $p_z$ orbitals of each carbon atom, and the $6 \times 2 \, \sigma$ bands, built from the $p_x , p_y$ and $s$ orbitals. These bands lie below and above the Dirac point defined by the $p_z$ orbitals.
\\ \indent
The resulting hamiltonian is then a matrix of dimensions $16 N_L \times 16  N_L$, where $N_L$ is the number of layers. The atomic spin-orbit hamiltonian is: ${\cal H}_{SOC}^{at} = \lambda_{at} \sum_n \vec{L}_n \vec{S}_n$,
where $n$ labels the atoms in the lattice, and $\vec{L}$ and $\vec{S}$ describe the orbital angular momentum and the spin of each atom. 
${\cal H}_{SOC}^{at}$ induces transitions between the $p_z$ orbital ($L=1$ and $L_z = 0$),  and the $p_x , p_y$ orbitals (combinations of $L=1$ and $L_z = \pm 1$ orbitals), accompanied by a spin flip. 
It also includes a diagonal term which breaks the degeneracy between $| p_\pm ,\uparrow \rangle$ and $| p_\pm, \downarrow \rangle$, where $p_{\pm} = p_x \pm i p_y$. 

\begin{figure}
\begin{center}
{\includegraphics[scale=0.25]{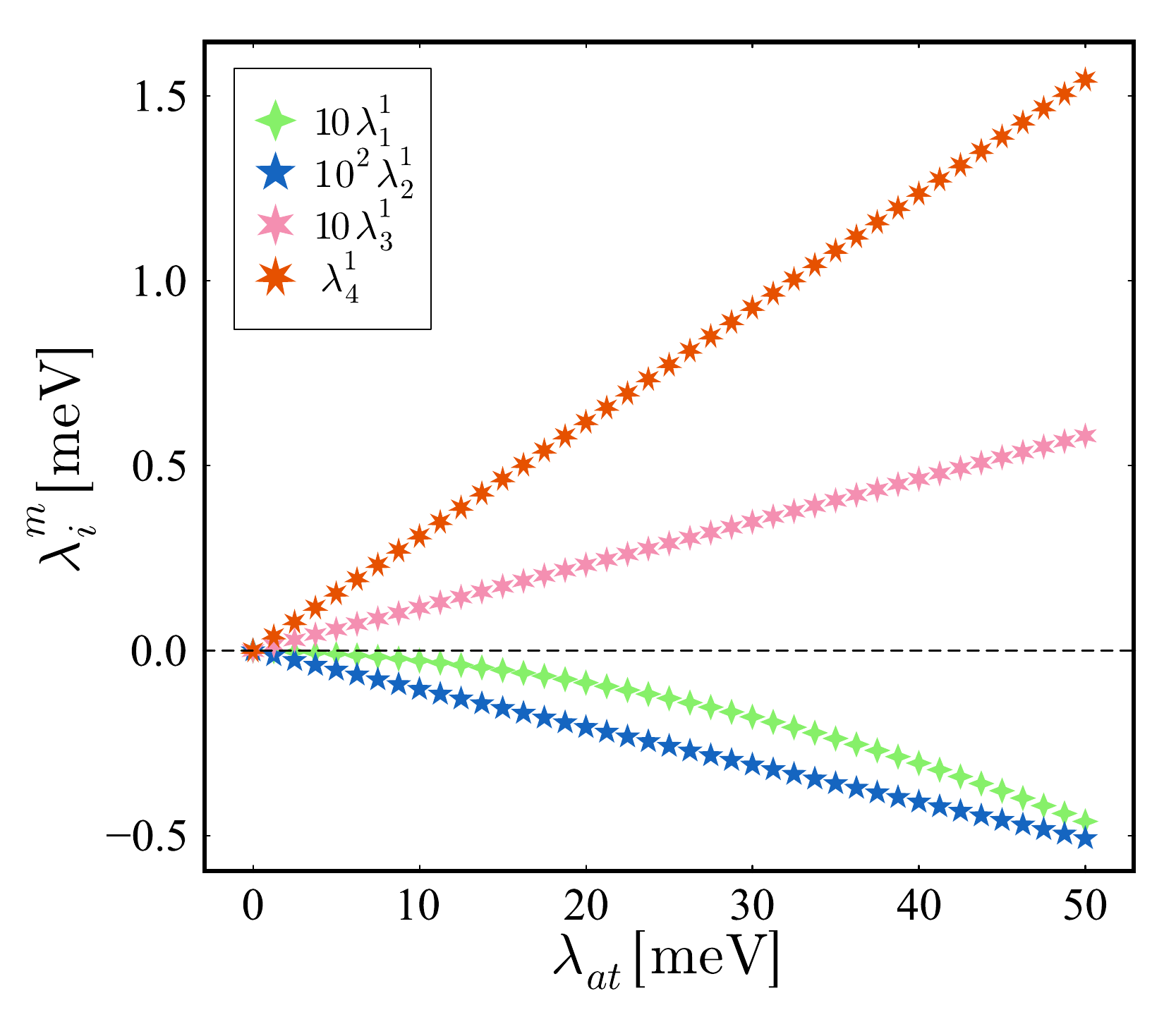}} 
\caption{
Scaling of the different effective $p_z - p_z$ spin orbit couplings terms $\lambda_1^1,\lambda_2^1,\lambda_3^1\lambda_4^1$ \, at the surface layer of an eight layer stack [see ${\cal H}_{SOC}$ in eq.~(\ref{hamilso_1})], as function of the value of the intra-atomic interaction, $\lambda_{at}$.
The calculation is carried out with the tight binding parameters shown in Table [\ref{tab:parameters}] of the SM~\cite{SM}, except for the parameter $\lambda_{at}$.} 

\label{fig:scaling_lambda}
\end{center}
\end{figure}

\emph{Projection of the spin-orbit coupling onto $p_z$ orbitals--}
As is commonly done for graphene, we define an effective low energy hamiltonian in terms of $p_z$ orbitals only. We consider the leading terms in a momentum expansion around the corners of the Brillouin zone (BZ).
By integrating out transitions which involve the other orbitals (see Section ~\ref{sec:sm-intout} of the SM~\cite{SM}) and analyzing the symmetries of a RMG stack, we find that the allowed terms for the SOC hamiltonian at the corners of the BZ are direct extensions of the intrinsic and Rashba terms in a graphene monolayer and in Bernal graphite~\cite{dresselhaus_spin-orbit_1965,guinea_spinorbit_2010,liu_giant_2010,mccann_spin-orbit_2010,konschuh_tight-binding_2010,konschuh_theory_2012}. 

In the following, we keep terms up to nearest neighbor layers and neglect the effect of a displacement field. Then, generalizing the interactions defined for Bernal graphite~\cite{dresselhaus_spin-orbit_1965} and a graphene bilayer~\cite{guinea_spinorbit_2010,konschuh_theory_2012}.

We obtain:
\begin{align}
    {\cal H}_{SOC} &=  \sum_{m=1}^{m=N_L}
    \left[ \lambda_1^{m}\tau_z \sigma_z^m s_z^m + \lambda_2^{m} ( -1 )^m \tau_z s_z^ m + \right. 
    \nonumber \\
    &+ \left.  \lambda_3^{m} ( - 1 )^m  \left(  \sigma_y^m s_x^m -  \tau_z \sigma_x^m s_y^m \right) \right] + 
    \nonumber \\
    &+  \sum_{m=1}^{m=N_L-1}
   \lambda_4^{m}   \left[ \sigma_z^m     \left(   \mu_y^{m,m+1}  s_x^m - \tau_z \mu_x^{m,m+1} s_y^m \right)  + h. c. \right]
    \label{hamilso_1}
\end{align}
where $\sigma^m , \tau^m$ and $s^m$ refer to the Pauli matrices which describe the sublattice, valley, and spin in a given layer. The Pauli matrix $\mu^{m,m+1}$ connects sites in neighboring layers leaving all other indices unchanged. 
$N_L$ is the number of layers. 
The parameters $\lambda_1^m , \lambda_2^m , \lambda_3^m$ and $\lambda_4^m$ are real and have dimensions of energy. 
Note that these terms depend on the layer index $m$. The dependence on $m$ of the parameters  $\lambda_1^m$ and $\lambda_4^m$ is very weak, while $\lambda_2^m$ and $\lambda_3^m$ decay to zero far from the surface, in agreement with the symmetries of bulk rhombohedral graphite, where only $\lambda_1$ and $\lambda_4$ are allowed. (see Section~\ref{sec:sm-lambdas} of the SM~\cite{SM}).

They can be obtained by directly solving the full hamiltonian, as described below, or by a perturbation expansion on $\lambda_{at}$. 
The dependence of these parameters at the surface layer on the value of the intra atomic coupling $\lambda_{at}$ is shown in Fig.~\ref{fig:scaling_lambda}. Results for the layer dependence in an eight layer stack are shown in Fig.~\ref{fig:lambdalayer} of the SM~\cite{SM}.
The Kane-Mele term, $\lambda_1$, scales quadratically on $\lambda_{at}$, as in a graphene monolayer, while the other parameters depend linearly~\cite{linear} on $\lambda_{at}$.
Within numerical accuracy, the results are independent of the number of layers and of the layer position. 

Away from $\vec{k} = 0$ we find a number of momentum dependent terms consistent with the symmetries of the system. 
In particular, we find a term:
\begin{align}
    \delta {\cal H}_{SOC}^{\vec{k}} &\sim s_z ( k_x \tau_z i k_y )^3
\end{align}
as in the surface of a topological insulator with trigonal warping~\cite{fu_hexagonal_2009}. 
This term leaves the sublattice, valley, and layer indices unchanged. 
It also does not modify significantly the results for fillings such that the Fermi level is within the flat surface bands.

\begin{table}[ht]
\centering\scriptsize
\setlength{\tabcolsep}{5.2pt}
\renewcommand{\arraystretch}{1.15}
\begin{tabular}{@{}l|ccccc@{}}
\hline
\hline
{\bf Phase} & $\Nb$ & $(N_b^K,N_b^{K'})$ & $\mathcal T$ & $|P_v|$ & $|P_{sv}|$  \\
\midrule
\midrule
F-R  & 4 & $\left( 2,2 \right)$ & \checkmark & $0$            & $\sim0$    \\
F-VI         & 4 & $\left( 2,2 \right)$ & \ding{55} & partial        & partial      \\
F-SVL        & 4 & $\left( 2,2 \right)$ & \checkmark & $0$            & partial          \\
\midrule
TQ-SVL       & 3 & $\left( 2,1 \right)$ & \ding{55} & $\sim1/3$ & $\sim1/3$            \\
TQ-R & 3 & $\left( 2,1 \right)$ & \ding{55} & partial        & $\sim0$           \\
\midrule
H-V          & 2 & $\left( 2,0 \right)$ & \ding{55} & $1$         & partial     \\
H-VI         & 2 & $\left( 1,1 \right)$ & \ding{55} & partial        & partial                \\
H-SVL        & 2 & $\left( 1,1 \right)$ & \checkmark & $0$            & $\sim1$         \\
H-R  & 2 & $\left( 1,1 \right)$ & \checkmark & $0$            & $\sim0$            \\
\midrule
Q-$z$        & 1 & $\left( 1,0 \right)$ & \ding{55} & $1$         & $\sim1$         \\
Q-R  & 1 & $\left( 1,0 \right)$ & \ding{55} & $1$         & $\sim0$     \\
\hline
\hline
\end{tabular}
\caption{Classification of the different states appearing in the phase diagram (Fig.~\ref{fig:master}a). 
The prefix denotes the number of doped bands as the sum of valley-resolved counts $N_b = N_b^{K} + N_b^{K'}$:
full (F), three-quarter (TQ), half (H), and quarter (Q) metal. 
The suffix is assigned from the valley polarization $P_v$ and the spin--valley locking polarization $P_{sv}$: 
valley imbalance (VI) for unequal valley populations with $0<|P_v|<1$; 
valley saturated (V) for $|P_v|=1$; 
spin--valley locked (SVL) for exchange-driven order with the carrier spins polarized along $\pm\hat{z}$ with opposite sign in the two valleys; 
and Rashba (R) for states with no spontaneous spin order, where the carriers spins follow an in-plane Rashba texture. 
In the quarter metal locking is undefined ($P_{sv}\equiv P_s$) and the spin-valley-polarized state is labeled $z$ instead. 
In the $|P_v|$ column,  $0$ and $1$ entries are exact, protected by time-reversal symmetry ($\mathcal{T}$)  and by valley saturation, respectively; ``partial'' denotes intermediate
values not pinned by symmetry, and the $\sim\!1/3$ entries reflect the natural $2\!:\!1$ ratio of the TQ count.}
\label{tab:catalogo}
\end{table}

\begin{figure*}[ht]
    \includegraphics[scale=0.315]{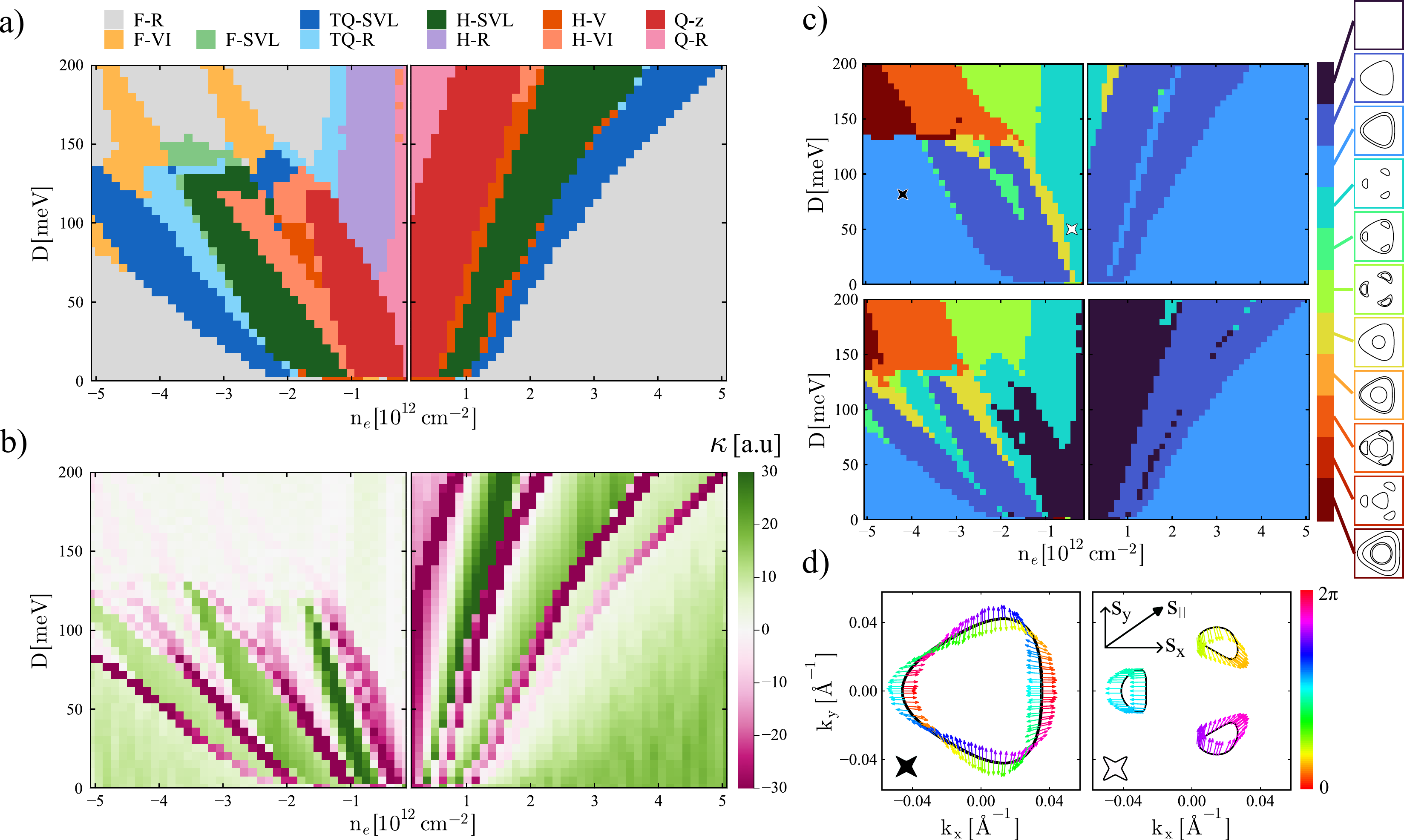}
    \caption{
    (a) Ground-state phase diagram as a function of external displacement field $D$ and carrier density $n_e$, for holes (left) and electrons (right); Color labels indicate the different phases detailed in Table~\ref{tab:catalogo}. 
    (b) Interaction inverse-compressibility $\kappa\sim d\mu/dn$. 
    (c) Valley-resolved Fermi surface topology index for the majority (top) and minority (bottom) valleys (see Sections ~\ref{sec:sm-bandpol} to ~\ref{sec:sm-fs} of the SM~\cite{SM} for definitions). 
    (d) Spin textures over the $K$-valley Fermi contours for TQ-SVL (left) and Q-R (right); $\mathbf{s}_\parallel$ is depicted by arrows whose color encodes the azimuthal angle. A winding phase emerges with opposed $\pm1$ helicity for each intra-valley doped band.
    } 
    \label{fig:master}
\end{figure*}

\emph{Perturbative expansion of some spin-orbit coupling terms--}
The parameter $\lambda_1$ is an extension of the Kane-Mele~\cite{kane_quantum_2005} term  of a single graphene layer.  In a single-layer, the spin-independent part of the Hamiltonian cannot have matrix elements between the $p_z$ and the $p_{\pm}$ and $s$ orbitals~\cite{kane_quantum_2005}. There are two sets of bands, the two $\pi$ bands based on the two $p_z$ orbitals in the unit cell, and  six $\sigma$ bands, based on the four $p_\pm$ and two $s$ orbitals. 
The effective SOC between $p_z$ orbitals requires two jumps of an $p_z$ electron to the $\sigma$ bands (see Fig.~\ref{fig:SOC_sketch}), induced by the atomic SOC, and it is quadratic on $\lambda_{at}$, so that it should scale as $\lambda_{at}^2$ divided by an energy scale comparable to the separation in energy of the $p_z$ band and the bands which include the $p_\pm$ and $s$ orbitals. 

A detailed expansion is presented in Section[\ref{sec:details}]. We obtain:
\begin{align} 
\lambda_1 &= \lambda_{at}^2 \frac{\epsilon_s}{18 t_{ps}^2}
\end{align}
This expression uses the definition of tight binding parameters in Section \ref{sec:tight_binding} of the SM~\cite{SM}.

The value obtained using the parameters in Table~\ref{tab:parameters} of the SM~\cite{SM} is $\lambda_1 \approx 6.5 \mu$eV, in reasonable agreement with previous estimates
\cite{kane_quantum_2005,huertas-hernando_spin-orbit_2006,min_intrinsic_2006,yao_spin-orbit_2007}. This result depends on intra layer parameters only, that it does not change with the number of layers.

As discussed below, the largest term found in the full calculations, is the interlayer term $\lambda_4$. 
This term involves a spin flip accompanied by a jump between nearest neighbor layers (see Fig.~\ref{fig:SOC_sketch}). 
This term can be obtained from first order perturbation theory in $\lambda_{at}$, as it is a combination of an intra-atomic transition induced by the atomic SOC, and a spin independent transition between a $p_{\pm}$ or an $s$ orbital in one layer and a $p_z$ in the next  (see Fig.~\ref{fig:SOC_sketch}). These transitions, which cannot exist in a single layer, are not forbidden by symmetry in the multilayer case. 
In Section~\ref{sec:details} of the SM~\cite{SM} we derive an analytical expression, based on the argument outlined above, which fits reasonably well the results from the full TB model over a significant range of parameters: 
\begin{align}
    \lambda_4 &\approx \lambda_{at}  \frac{\epsilon_s ( t_\pi^{il,nn} +  t_\sigma^{il,nn} )}{9 t_{ps}^2}
\end{align}
Using the parameters in Table~\ref{tab:parameters} of the SM~\cite{SM} we obtain $\lambda_4 \approx 0.61$ meV. Note that the Fermi velocity is $v_F = ( t_\sigma d ) / ( 2 \sqrt{3} )$.

As a final note, we would like to highlight that the problem of graphene stacks with SOC can be mapped to a Su-Schrieffer-Heeger hamiltonian, which we develop in Section~\ref{sec:simplified} of the SM~\cite{SM}. 

\emph{Mean-field phase diagram: Spin textures in rhombohedral multilayer graphene--}
Having established $\lambda_4$ as the intrinsic SOC leading term in RMG, we now include correlations in our calculation in order to study the competing phases in R4G. 
To do so, we use mean-field Hartree-Fock (HF) theory using a continuum model incorporating intrinsic SOC (see Sections ~\ref{sec:sm-contmod} and ~\ref{sec:sm-hfmod} in the SM~\cite{SM}).

Throughout this work we consider only states that preserve valley $U(1)_v$ symmetry (hence, there are no inter-valley coherent phases), so that we can use the valley polarization $P_v=\langle\tau_z\rangle$ and
the valley-band-resolved counting $(N_b^{K},N_b^{K'})$ as good quantum numbers.
Since $\lambda_4$ couples spin to the layer and sublattice degrees of freedom through the in-plane spin operators $\hat S_x$ and $\hat S_y$, it breaks $SU(2)$ spin symmetry explicitly, so that $P_s=\langle s_z\rangle$ is no longer well defined.  
To fully identify the ground states of the system, we need an extra order parameter: the spin--valley polarization $P_{sv}=\langle\tau_z s_z\rangle$, which measures the collinear locking of the out-of-plane spin to the valley.
The different ground states are classified in Table~\ref{tab:catalogo} (see also Sections ~\ref{sec:sm-bandpol} and ~\ref{sec:sm-valleypol} of the SM~\cite{SM}).

The interacting phase diagram is shown in Fig.~\ref{fig:master}(a).
Doping away from charge neutrality, the interacting ground states organize into a cascade of flavor-symmetry-broken metals whose sequence differs sharply between holes and electrons. 
We trace this asymmetry due to the stronger effect of the exchange term on the conduction bands (due to the flat character of the band), which orders the carriers before the splitting due to $\lambda_4$ can energetically compete.
The negative inverse compressibility $\kappa=\partial \mu/\partial n$ [Fig.~\ref{fig:master}(b)], is consistent with the first-order character of these transitions between states in the phase diagram. 

For electron doping the system transitions from a quarter-metal ($N_b=1$) to a full-metal ($N_b=4$) smoothly by changing the doping. Here, exchange correlations dominate the leading energetics and half-, and three-quarter- states with a Rashba (-R) spin-texture never condense.
On the contrary, on the hole side the evolution is less uniform and features competing domes. 
Among this plethora of domains, we find an interesting half-metallic-Rashba (H-R) state. 
Although it shares the $(1,1)$ band count and time-reversal symmetry ($\mathcal{T}$) invariance of the 
H-SVL, their origin is opposite. 
In the region where H-R condenses, the strength of the $\lambda_4$ term is energetically more favorable than that of exchange;
however, on the electron side, where a flat band emerges, $\lambda_4$ is not large enough to win this competition, and only H-SVL is present.
H-R is, therefore, a direct manifestation of the importance of intrinsic interlayer SOC in graphene multilayers.

As the $\lambda_4$ term induces a band splitting, there is a wide catalog of valley-resolved FSs.
We characterize all the possible FS topologies [see Fig.~\ref{fig:master}(c)] by a single index that combines the Euler characteristic, $\chi$, of the carrier region, with the band count (see Section ~\ref{sec:sm-fs} of the SM~\cite{SM}). 
Depending on doping and displacement field, the FS changes both shape and topology, becoming circular, annular, pocket-like, or a combination of these. 
For electron doping, the change of topology of the FS follows mainly the phase changes. 
Nevertheless, we highlight the low density and high displacement field transition from 3-pockets to annular within the Q-R phase, with a van-hove singularity where superconductivity has been previously studied~\cite{parra-martinez_band_2025}. 
In the hole side, the interplay between trigonal warping ($\gamma_3$, see the continuum model described in Section~\ref{sec:sm-contmod} of the SM~\cite{SM}) --which enhances the fragmentation of the FS into different pockets characteristic of RMG--, and the intrinsic SOC, realizes a broad set of topologies where the band splitting by $\sim\!2\lambda_4$ gains relevance. 
It is important to note that fermiology and flavor order are independent degrees of freedom as different FS shapes appear within a specific state.

Since the  $\lambda_4$ term is related with $S_x$ and $S_y$, carriers develop an in-plane spin texture [see Fig.~\ref{fig:master}(d)]. As we can see in Fig.~\ref{fig:master}(d), when the two intra-valley bands are doped (e.g. TQ-SVL state), they wind with opposite $\pm1$ helicity. 
Nevertheless, for the Q-R state, even though just a single band per valley is occupied, it still shows a winding texture, with its opposite-helicity partner unoccupied due to the intrinsic SOC band splitting. 
The in-plane magnetization is always zero since the $2\pi$ winding cancels out when integrating over the BZ.
 
\begin{figure}[t]
\begin{center}
{\includegraphics[scale=0.25]{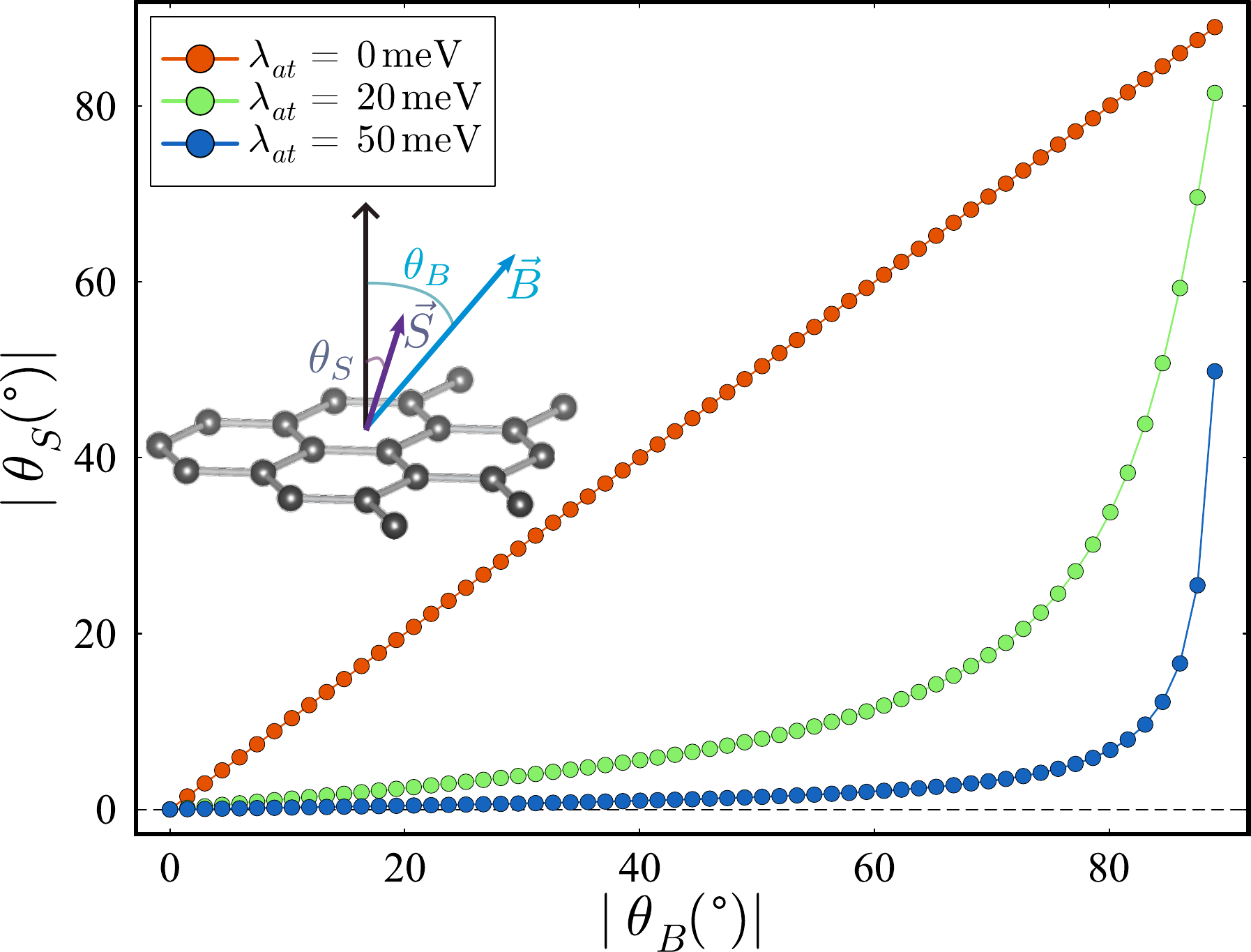}} 
\caption{
Tilting of a spin under an applied out of plane magnetic field, $B_\perp = 0.015 $ T and a varying in plane magnetic field, $B_\parallel$. The angles are $\theta_B = \arctan ( B_\parallel / B_\perp )$, and $\theta_S = \arctan ( m_\parallel / m_\perp )$, where $m_\parallel , m_\perp$ label the induced magnetization. 
These angles are calculated for different values of the intraatomic spin-orbit coupling, $\lambda_{at}$, see~\cite{auerbach_isospin_2025}. Inset: Sketch of the directions of the magnetic field and spin with respect to the last layer of graphene.
}
\label{fig:tilted}
\end{center}
\end{figure}

\emph{Effect of an in plane magnetic field--}
The spin-orbit coupling in rhombohedral graphene has been studied experimentally by measuring the magnetization induced by an in-plane magnetic field, which has only a Zeeman coupling to the electron spin~\cite{arp_intervalley_2024,auerbach_isospin_2025}. 
It was generally assumed that a Kane-Mele or an Ising coupling in rhombohedral graphene lead to the pinning of the spin in the out-of-plane direction. 

In the following we consider the competition between the $\lambda_1$ and $\lambda_4$ terms in eq.~(\ref{hamilso_1}), which favor an out-of-plane and an in-plane alignment, respectively. 
The interlayer spin independent hopping is described by a term which connects atoms in nearest layers aligned along the vertical direction, $t_\perp$~\cite{footnote}. 
The spin Hamiltonian projected onto the two spin flat bands at one surface, in the presence of a in-plane magnetic field, $B_\parallel$ is: 
\begin{align}
    {\cal H}_{SOC} ( \vec{k} ) &= \lambda_1 \tau_z \sigma_z  s_z + 
    \frac{\lambda_4 v_F \tau_z\sigma_0}{t_\perp}  [ ( k_x + i k_y )   s_- + \nonumber \\ 
    & + ( k_x - i k_y ) s_+ ] + B_x \mu_B \tau_0 \sigma_0 s_x
    \label{eq:hamil_magf}
\end{align}
where $\mu_B$ is the Zeeman coupling, and we assume that the applied field is along the $x$ direction.

The value of the observed magnetization depends on the values of $\lambda_1 , \lambda_4$, $\mu_B B_0$, and the carrier density, $n$. Results form the full TB model are shown in
Fig.~\ref{fig:tilted}. A simplified model using the flat bands only, and assumming that $\lambda_4 \gg \mu_B B_0 , \lambda_1$ gives the expression  (see Section~\ref{sec:tilted} of the SM~\cite{SM}):
\begin{align}
    \theta_S &= \arctan \left[ \frac{\mu_B B_0 \tan ( \theta_B )}{2 ( \lambda_1 + \mu_B B_0)} \right]
    \label{eq:til2}
\end{align}
This expression is independent of $\lambda_4$. In the limit $\mu_B B_0 \gg \lambda_1$ we get $\tan{\theta_s}=\tan\frac{\theta_B}{2}$. 
Consequently, at $\theta_B=0,\pi/2$ the magnetization follows the external magnetic field, as expected by symmetry, but away from these values, and in particular close to $\theta_B=\pi/2$, the magnetization tends to stay out of plane even when the in-plane field is strong, mimicking a large Ising-like SOC \cite{Zeldov_private_comm}. 


\emph{Conclusions--}
We have shown that intrinsic spin-orbit coupling in rhombohedral graphene stacks is dominated by an interlayer term, $\lambda_4$, that combines an atomic spin-flip transition with a nearest-layer hopping process. 
Unlike the Kane-Mele-like term $\lambda_1$ --which survives in a single-layer and scales quadratically with the atomic coupling $\lambda_{at}$-- $\lambda_{4}$ is forbidden by symmetry in the monolayer case and scales linearly with the atomic $\vec{L} \vec{S}$ coupling, $\lambda_{at}$, making it around two orders of magnitude larger than the Kane-Mele term, $\lambda_1$. 

Including $\lambda_4$ in a mean-field Hartree-Fock calculations of the specific case of R4G reveals that this interlayer SOC has qualitative consequences for the correlated phase diagram. 
It couples spin to the layer and sublattice degrees of freedom, explicitly breaking spin-rotation symmetry.
This produces a richer catalog of flavor-symmetry-broken metallic phases and FSs than those found in calculations without SOC or even the one expected from a Kane-Mele-type coupling alone. 

Finally, we identified an experimentally accessible way to distinguish which SOC mechanism dominates in a given sample: the magnetization induced by a weak in-plane magnetic field scales linearly with carrier density if $\lambda_1$ is dominant, but as the square root of density if $\lambda_4$ prevails. 
This distinct density dependence, together with the different states found in the phase diagram, can be used as signatures in future magnetotransport and magnetometry experiments on rhombohedral graphene multilayers.

More broadly, our Letter indicates that graphitic multilayers should not be treated as simple extensions of single-layer graphene when it comes to spin-orbit physics: stacking opens interlayer terms in the Hamiltonian that are symmetry-forbidden in a monolayer and can dominate the effective SOC by more than an order of magnitude. 
We expect this interlayer mechanism, and its consequences for correlated order and Fermi surface topology, to be relevant across the growing family of rhombohedral and twisted graphitic multilayers. 

\emph{Acknowledgments--}
We are thankful to E. Zeldov for relevant comments about this manuscript.
IMDEA Nanociencia acknowledges support from the Spanish State Research Agency (AEI/10.13039/501100011033) and the Ministry of Science, Innovation and Universities (MICIU) through the `Severo Ochoa" Programme for Centres of Excellence in R\&D (CEX2020-001039-S and CEX2025-001660-S). 
We acknowledge support from NOVMOMAT, project PID2022-142162NB-I00 funded by MICIU/AEI/10.13039/501100011033 and by FEDER, UE.
J.A. S.-G. has received financial support through the ``Ram\'on y Cajal'' Fellowship program, grant RYC2023-044383-I financed by MICIU/AEI/10.13039/501100011033 and FSE+.
G.P.-M. is supported by Comunidad de Madrid through the PIPF2022 programme (grant number PIPF-2022TEC-26326).   
A.J.-P. and F.G. acknowledge funding from the EU NextGenerationEU/PRTR-C17.I1, the Basque Government’s Department of Education through project PIBA\_2023\_1\_0007 (STRAINER), as well as by the IKUR Strategy under the collaboration agreement between Ikerbasque Foundation and DIPC on behalf of the Department of Education of the Basque Government.
We thankfully acknowledge RES resources provided by BSC in MareNostrum5 to FI-2025-2-0024, FI-2025-3-0040 and FI-2026-1-0063. 
AS acknowledges support by  the Minerva Stiftung, the
DFG through CRC/Transregio 183 (Project Grant No.
277101999), the
Israel Science Foundation ISF (Grant No 1914/24), as
well as ISF Quantum Science and Technology (2478/24). 
Claude (Sonnet5 and Opus5) has been used to improve the efficiency of the numerical codes and certain sections of the text.

\emph{Data availability--}
The data that support the findings of this article are not publicly available. 
The data are available from the authors upon reasonable request.

\let\oldaddcontentsline\addcontentsline 
\renewcommand{\addcontentsline}[3]{}    
\bibliography{rh_soc_final}

\begin{thebibliography}{60}
\bibitem [{foo()}]{footnote}%
  \BibitemOpen
  \href@noop {} {}\bibinfo {note} {It is worth noting that keeping the $\lambda_4$ term and the inter layer hopping between nearest neighbors only, and projecting the resulting hamiltonian onto the flat bands of a single surface, a simple Dirac hamiltonian is obtained, see \ref{sec:simplified}.}\BibitemShut {Stop}%
\end{thebibliography}
\let\addcontentsline\oldaddcontentsline 

\clearpage
\ %

\onecolumngrid
\newpage
\thispagestyle{empty}
\setcounter{equation}{0}
\setcounter{figure}{0}
\setcounter{table}{0}
\setcounter{page}{0}
\setcounter{section}{0}

\renewcommand{\theequation}{S\arabic{equation}}
\renewcommand{\thefigure}{S\arabic{figure}}
\renewcommand{\thesection}{S\arabic{section}}
\renewcommand{\thetable}{S\arabic{table}}

\begin{center}
\Large Supplemental Material for \\
Intrinsic spin-orbit coupling in graphene stacks. \\
\end{center}

\begin{center}
\normalsize Guillermo Parra-Mart\'inez, Alejandro Jimeno-Pozo, Ady Stern, Jos\'e \'Angel Silva-Guill\'en and Francisco Guinea
\end{center}

\tableofcontents
\clearpage

\makeatletter
\def\@hangfrom@section#1#2#3{\@hangfrom{#1#2}#3}
\def\@hangfroms@section#1#2{#1#2}
\makeatother
\section{Tight-binding description of multilayer graphene, its parameters and the resulting band structure}
\label{sec:tight_binding}
\begin{figure*}[h!]
    \includegraphics[scale=0.95]{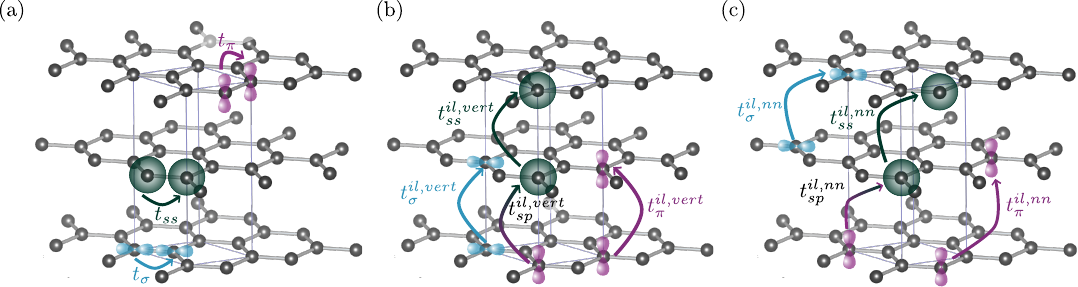}
    \caption{Schematics of the TB parameters of the model. (a) Intralayer hoppings. (b) Vertical interlayer hoppings. (c) Nearest-neighbours interlayer hoppings.}
    \label{fig:hoppings}
\end{figure*}

The spin independent part of the tight-binding (TB) hamiltonian requires the hoppings between $p_i$ orbitals at different sites, between $s$ orbitals, and between $p_i$ and $s$ orbitals. For each pair of sites, the $p_i - p_j$ hoppings can be characterized in terms of Slater-Koster parameters $t_\pi$ and $t_\sigma$, and two other parameters, $t_{sp}$ and $t_{ss}$ need to be defined. Finally, there is an onsite term, $\epsilon_{s}$, and we take $\epsilon_p = 0$ (this parameter defines the Dirac energy in a monolayer). We restrict the hoppings to nearest neighbor within each layer and between nearest neighboring layers. With this restriction, four parameters are required to describe the intra-layer hoppings, four parameters for the hoppings between atoms in different layers right on top of each other, and eight parameters for the hoppings between atoms which are first nearest neighbors in different layers, which are shown schematically in Fig.~\ref{fig:hoppings}. This choice of parameters follows~\cite{tomanek_first-principles_1988,tomanek_growth_1991}. The parameter which can be known with the highest accuracy is $V_\sigma$ between orbitals in the same layer, as it determines the Fermi velocity at the Dirac point of the $p_z$ band. Hoppings between $p_{\pm}$ and $s$ orbitals in different layers are poorly known, although they are expected to be in the range $0.1 - 0.4$ eV, as they are related to the inter-layer $p_z - p_z$ terms $\gamma_3$ and $\gamma_4$ in Bernal graphite. 
The parameters used in the calculations reported below are listed in Table~\ref{tab:parameters}. Note that, if one uses a Slater-Koster convention for the dependence on distance and orientation of the hoppings between $p$ orbitals, the in- and out-of-plane hoppings are related.

\begin{table*}[h!]
\begin{tabular}{cccccccccccccc}
\hline
\hline

$\lambda_{at}$ & $t_\pi$  & $t_\sigma$ & $t_{ps}$ & $t_{ss}$ & $\epsilon_s$ & $t_\pi^{il, vert}$ & 
$t_\sigma^{il, vert}$ & $t_{sp}^{il, vert}$ & $t_{ss}^{il, vert}$ & $t_\pi^{il , nn}$ & 
 $t_\sigma^{il , nn}$ & $t_{sp}^{il , nn}$ & $t_{ss}^{il , nn}$ \\

0.02  & 6.38 & 2.66 & 4.98 & 4.3 & -7.3 & 0.4 & 0.4 & 0.53 & 4.3 & 0.4 (0.73) & 0.4 (0.3)& 0.2 & 0 \\

\hline \hline
\end{tabular}
\caption{TB parameters used in the model. The super index $il$ means coupling between sites in  neighboring layers, $vert$ describes atoms directly on top of each other, and $nn$ describes nearest neighbor atoms in neighboring layers. Values in eV. See\cite{tomanek_first-principles_1988,tomanek_growth_1991}. The hoppings between $p$ orbitals in neighboring layers which are not directly on top of each other are given by $t_\pi^{il , nn}$ and $t_\sigma^{il , nn}$ multiplied by factors which depend on their orientation and distance. The values in parenthesis give the final hoppings between a $p_z$ orbital in one layer and a $p_{x,y}$ orbital in another layer.  }
\label{tab:parameters}
\end{table*}

\begin{figure}
    \includegraphics[scale=0.3]{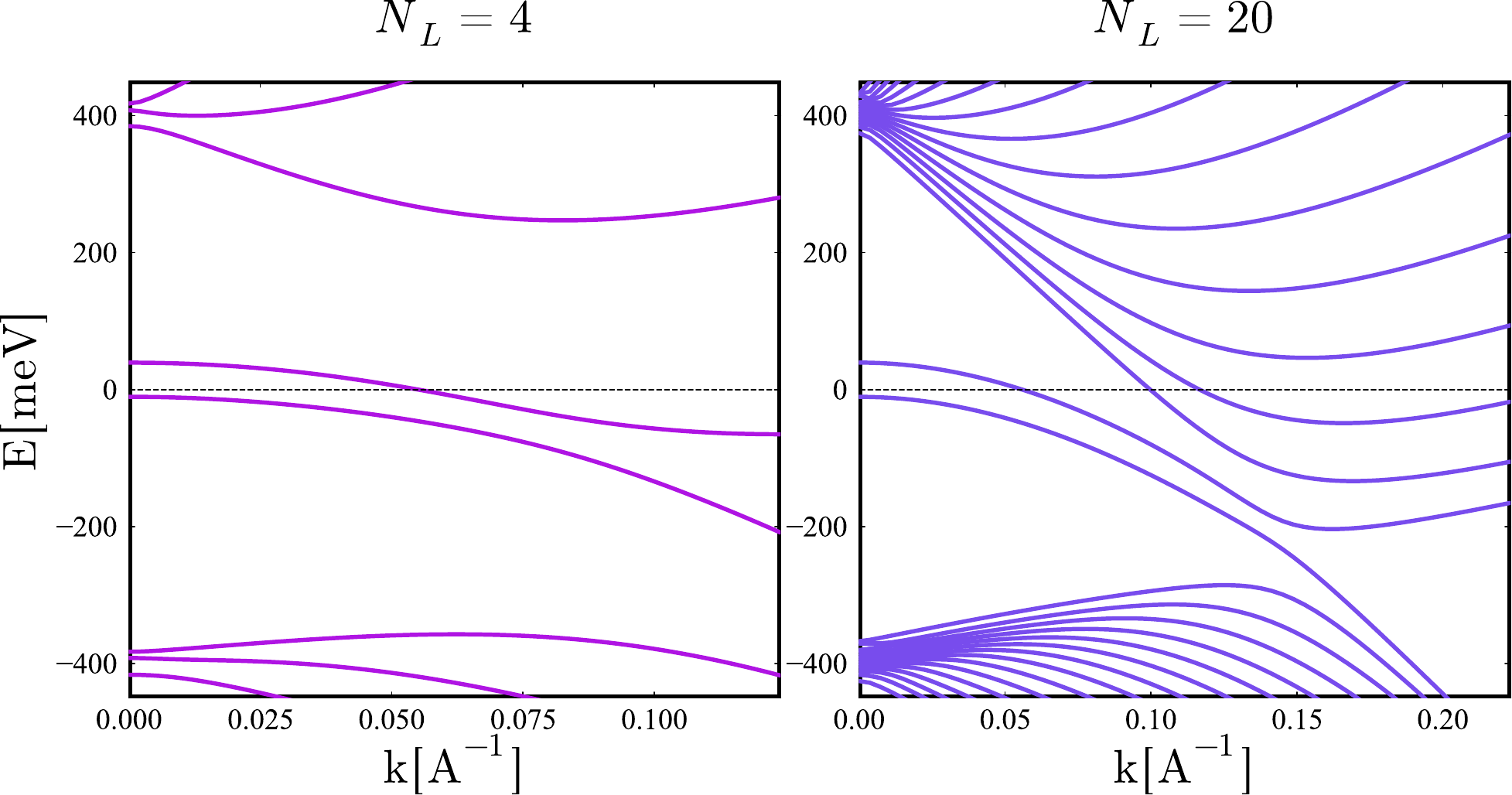}
    \caption{Tight-binding electronic bands along the $y$ direction, starting at the point $\vec{K} = [ ( 4 \pi )/ (3 a ) , 0 ] $, where $a \approx 1.42 $ \AA \, is the inter-atomic distance. The bands near E $\approx  0$ form two doublets, whose separation, due to the spin-orbit coupling is not visible. The momentum scale $k_c =\gamma_1 / v_F$ is $k_c \approx 0.12$ \AA$^{-1}$. 4 layer stack (left) and 20 layer stack (right).}
    \label{fig:tbbands}
\end{figure}

\section{Integrating out the $s$ and $p_x,p_y$ orbitals for the effective model}\label{sec:sm-intout}
The effective hamiltonian can be written schematically as:
\begin{align}
    {\cal H}_{p_z , p_z}^{eff} &\approx {\cal H}_{p_z , p_z} + \sum_{i = p_x , p_ y , s} {\cal H}_{p_z , i} {\cal H}_{i , j}^{-1} {\cal H}_{j , p_z}  
\end{align}
where ${\cal H}$${_{i,j}}$ is the full hamiltonian. 
The spin orbit coupling is then:
\begin{align}
    {\cal H}_{SOC} &= \left. {\cal H}_{p_z , p_z}^{eff} \right|_{\lambda_{at}} - \left. {\cal H}_{p_z , p_z}^{eff} \right|_{\lambda_{at} = 0}
\end{align} 

\clearpage

\section{Layer dependence of $\lambda_i^m$}
\label{sec:sm-lambdas}

\begin{figure*}[h]
\begin{center}
\begin{tabular}{c}
{\includegraphics[width=\textwidth]{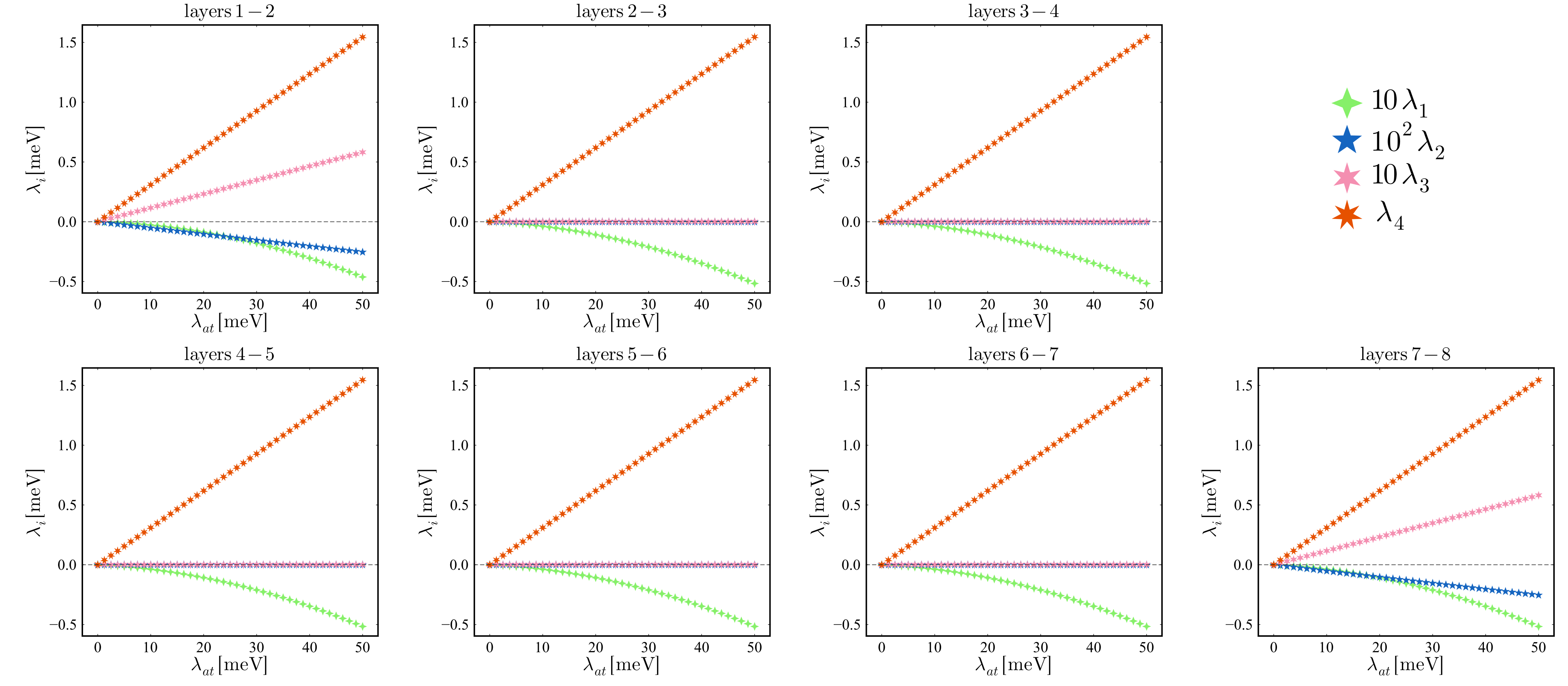}} 
\end{tabular}
\caption{Layer dependence of $\lambda_i$ in an eight layer stack for different pairs of neighboring layers.}
\label{fig:lambdalayer}
\end{center}
\end{figure*}

Fig.[\ref{fig:lambdalayer}] shows the layer dependence of the parameters $\lambda_1^m , \lambda_2^m , \lambda_3^m , \lambda_4^m$ for an eight layer stack. Note that the parameters $\lambda_3^m$ and $\lambda_4^m$ involve layers $m$ and $m + 1$. At the center of the stack the parameters $\lambda_2^m$ and $\lambda_3^m$ are negligible, in agreement with the symmetries of an infinite rhombohedral stack.

\clearpage

\section{Details of the perturbative expansion of the spin-orbit coupling}
\label{sec:details}
The tight binding model described previously can be expanded in powers of $\lambda_{at}$ in order to obtain analytical expressions for the spin orbit coupling in the $p_z$ bands.
\\
\indent
In a monolayer, the hybridization between $p_{\pm}$ and $s$ orbitals leads to the valence and conduction bands. Using the nearest neighbor model described above, with four orbitals per site, the $s$ orbital in one sublattice and the $p_+$ in the other sublattice (or $p_-$, depending on valley and sublattice) define a separate reduced $2 \times 2$ hamiltonian made from $| p _\pm \rangle$ orbitals in one atom, and $\ s \rangle$ orbitals in the neighboring ones. In terms of the parameters in Table~\ref{tab:parameters}, this hamiltonian is:
\begin{align}
{\cal H}_{red}^{p_\pm , s} &= \left( \begin{array}{lr} 0 &3   t_{ps} / \sqrt{2} \\3   t_{ps} / \sqrt{2} &\epsilon_s
\end{array}
\right)
\end{align}
The perturbatively term can be obtained by assuming a transition, induced by ${\cal H}_{SOC}^{at}$, from the $p_z$ band to the $p_{\pm} , s$ bands, followed by another transition back to the $p_z$ band. 
The spin orbit projected onto the $p_z$ band is:
\begin{align}
{\cal H}_{SOC}^{\lambda_1} &= {\cal H}_{SOC}^{at} \left( \left.{\cal H}_{red}^{p_\pm , s} \right)^{-1} \right|_{p_\pm,p_\pm} {\cal H}_{SOC}^{at}
    \end{align}
This analysis leads to (note that ${\cal H}_{SOC}^{at}$ adds a factor $\lambda_{at}^2 / 2$ ):
\begin{align} 
\lambda_1 &= \lambda_{at}^2 \frac{\epsilon_s}{18 t_{sp}^2}
\end{align}
This expression, calculated for a monolayer, gives also the leading term in rhombohedral multilayers, as it depends only on intra layer hopping processes, irrespective of the number of layers. Near a surface, there is another term, linear in $\lambda_{at}$, which contributes to the value of $\lambda_1$. It involves the sequence of hoppings:
\begin{align}
    | A ,l , p_z . \uparrow &\rangle \rightarrow | A ,l + 1 , p_+ , \uparrow \rangle \rightarrow  | A , l , p_z , \uparrow \rangle
    \nonumber \\
     | B ,l , p_z . \downarrow &\rangle \rightarrow | B ,l - 1 , p_- , \downarrow \rangle \rightarrow  | B , l , p_z , \downarrow \rangle
\end{align}
where the dependence on the spin-orbit coupling arises from the dependence of the energy of the intermediate state on $L_z s_z$. The contribution of these processes to the value of $\lambda_1$ is of order  $\approx [ ( \epsilon_s ( t^{il,nn}_\sigma + t^{il,nn}_\pi ) / ( 3 t_{ps}^2 ) ]^2 \lambda_{at} $, where $l$ is a layer index. This contribution is smaller than the intra-layer contribution to $\lambda_1$ for $\lambda_{at} \gtrsim 5 \mu$eV.

\indent
The interlayer spin orbit coupling $\lambda_4$ involves an intra atomic spin flip hopping, $| p_{z \uparrow} \rangle \rightarrow | p_{A 1 \pm \downarrow} \rangle$, followed by a spin independent inter layer hopping, $| p_{ \pm \downarrow} \rangle \rightarrow | p_{A 2 z \downarrow} \rangle$, where the subindex $Ai ,  i=1,2$ labels sublattice and layer.
The contribution of these processes to the spin orbit coupling is;
\begin{align}
{\cal H}_{SOC}^{\lambda_4} &= {\cal H}_{SOC}^{at}  \left.{\cal H}_{red}^{p_\pm , s}  \right|_{p_\pm,p_\pm}^{-1} {\cal H}_{p_\pm , p_z}
    \end{align}
    where the last term is the spin independent part of the hamiltonian which connects a $p_\pm$ orbital in one layer to a $p_z$ orbital in the same sublattice in a neighboring layer. This contribution is approximately given by ${\cal H}_{p_\pm , p_z} \approx 3 \sqrt{2} ( t_\pi^{il,nn} +  t_\sigma^{il,nn} ) $ so that:
\begin{align}
    \lambda_4 &\approx \frac{ \lambda_{at}}{9} \frac{\epsilon_s ( t_\pi^{il,nn} +  t_\sigma^{il,nn} )}{ t_{sp}^2}
\end{align}
This expression reproduces accurately the value of $\lambda_4$ calculated numerically using the full tight binding model over a broad range of parameters.

\clearpage

\section{Non-interacting continuum model}\label{sec:sm-contmod}

We model the band structure of rhombohedral multilayer graphene for $N_L$ layers
via a conventional hopping parametrization, where the 
non-interacting part of the
Hamiltonian, which contains the displacement field and SOC terms, in the sublattice ($\alpha=A,B$) $\otimes$ layer
($l=1,2,\dots,N_L$) basis $S_0=\{A_1,B_1,\dots,A_{N_L},B_{N_L}\}$ is given by: 

\begin{gather}
\label{eq:main_hamil}
    \mathbb{K}_{\alpha l,\alpha' l'}(\mathbf{k}) = \begin{pmatrix}
\mathcal{L} (\mathbf{k})& \mathcal{V}(\mathbf{k}) & \mathcal{W} & \mathbb{0} & \cdots & \mathbb{0} \\
 & \mathcal{L}(\mathbf{k}) & \mathcal{V}(\mathbf{k}) & \mathcal{W} & \ddots & \mathbb{0} \\
 &  &  & \ddots & \ddots & \mathcal{W} \\
 &  &  &  & \ddots & \mathcal{V}(\mathbf{k}) \\
 &  &  &  &  & \mathcal{L}(\mathbf{k})
\end{pmatrix}  
+ \delta\begin{pmatrix}
    (\mathbb{I}-\sigma_z)/2 & \mathbb{0} & \cdots & \mathbb{0}\\
    \mathbb{0} & \mathbb{I} &  \cdots & \mathbb{0}\\
    \vdots & \vdots &  \ddots & \mathbb{0}\\
    \mathbb{0} & \mathbb{0} &  \mathbb{0} & (\mathbb{I}+\sigma_z)/2\\
\end{pmatrix}\\
+ \Delta\begin{pmatrix}
    (N_L-1)\mathbb{I} & \mathbb{0} & \cdots & \mathbb{0}\\
    \mathbb{0} & (N_L-3)\mathbb{I} &  \cdots & \mathbb{0}\\
    \vdots & \vdots &  \ddots & \mathbb{0}\\
    \mathbb{0} & \mathbb{0} &  \mathbb{0} & -(N_L-1)\mathbb{I}\nonumber\\
\end{pmatrix} +H_{SOC}^{(\lambda_4)}\nonumber
\end{gather}
where
\begin{subequations}
\begin{equation}
    \mathcal{L}(\mathbf{k}) = \mqty(0 & v_{0}\pi^{\dagger}(\mathbf{k}) \\ v_{0}\pi(\mathbf{k}) & 0), \quad
    \mathcal{V}(\mathbf{k}) = \mqty(v_{4}\pi^{\dagger}(\mathbf{k}) & v_{3}\pi(\mathbf{k}) \\ \gamma_{1} & v_{4}\pi^{\dagger}(\mathbf{k})), \quad
    \mathcal{W} = \mqty(0 & \gamma_{2}/2 \\ 0 & 0)
\end{equation}
\end{subequations}
with $\pi(\mathbf{k})=\xi k_x+ik_y$ and $v_{i}=\frac{\sqrt{3}}{2}a\gamma_{i}$,
being $a=2.46$~\AA{} the graphene lattice constant and $\gamma_i$ the hoppings. The kinetic energy is given by $\mathbb{K}$, $\delta$ is a sublattice potential, and $\Delta$ is an external electric field.
The momentum $\mathbf{k}$ is measured from the Dirac point of valley
$\xi=\pm1$, and the two valleys are related by time reversal. The parameter
$\delta$ accounts for the on-site energy difference of the dimerized sites. 

The leading spin-orbit term reads:
\begin{gather}
\label{eq:hsoc}
\mathcal{H}_{SOC}^{\lambda_4}=\sum_{m=1}^{N_{L}-1}\lambda_{4}\,\sigma_{z}^{m}
\big(\mu_{y}^{m,m+1}s_{x}^{m}-\tau_{z}\,\mu_{x}^{m,m+1}s_{y}^{m}\big)
\end{gather}
where $s_i$, $\sigma_i$ and $\tau_i$ are the Pauli matrices in spin, sublattice
and valley space, respectively, $\xi=\pm1$ denotes the eigenvalue of $\tau_z$, and
$\mu_{i}^{m,m+1}$ acts in the two-dimensional space spanned by the layers $m$
and $m+1$, so that the term is an interlayer process connecting adjacent
layers. In matrix block form, the coupling between the layers $m$ and $m+1$
can be written compactly as:
\begin{equation}
\Big[\mathcal{H}_{SOC}^{\lambda_4}\Big]_{m,m+1}
=-i\lambda_4\;s^{\xi}\,\sigma_z+\mathrm{h.c.} ,
\qquad s^{\xi}\equiv s_x-i\,\xi\,s_y
\label{eq:compacta}
\end{equation}
This term is essentially an interlayer hopping that flips the spin, raising it in one valley
and lowering it in the other, diagonal in sublattice with a $\sigma_z$ sign and
independent of $\mathbf{k}$. The momentum dependence of the resulting spin
textures is therefore inherited from the orbital wavefunctions rather than from
the coupling itself. Its symmetry content determines the structure of the
whole phase diagram. The term preserves the valley $U(1)_\xi$ symmetry so that the Hamiltonian remains block diagonal in $\xi$, which allows the valley-resolved diagonalization and makes the band counting $\Nb^{\xi}$ well defined. In contrast, it breaks the spin rotation symmetry
explicitly: because $s_x$ and $s_y$ enter with phases locked to the lattice
and the interlayer hopping does not conserve the layer index, no continuous
spin rotation survives. 

In the calculations, the continuum model is defined on a hexagonal grid
centered at the Dirac point up to a momentum cut-off $q_{c}=0.07\,K_{D}$,
where $K_{D}=\frac{4\pi}{3a}(1,0)$ is the position of the Dirac point,
corresponding to one corner of the Brillouin zone. The numerical values used
for the continuum model parameters are listed in Table~\ref{tab:kpparameters} and
taken from existing
literature~\cite{zhou_half_2021, Zibrov2018GullyTG}.

\begin{table*}[h!]
\begin{tabular}{cccccc}
\hline
\hline

$\gamma_0$ & $\gamma_1$ & $\gamma_2$ & $\gamma_3$ & $\gamma_4$ &
$\delta$ \\

3.1 & 0.38 & $-0.015$ & $-0.29$ & $-0.141$ & 0.0105  \\

\hline \hline
\end{tabular}
\caption{Continuum model parameters. $\gamma_0$ is the intralayer nearest-neighbor
hopping, $\gamma_1$ the interlayer coupling between dimerized sites, $\gamma_2$
the coupling between next-nearest layers, $\gamma_3$ the trigonal warping term
and $\gamma_4$ the electron--hole asymmetry term. $\delta$ is the on-site energy
of the dimerized sites. All values are displayed in eV.}
\label{tab:kpparameters}
\end{table*}

\clearpage
\section{Hartree-Fock approach for RMG}\label{sec:sm-hfmod}

In the following we define the interacting Hamiltonian within the
self-consistent Hartree--Fock (HF) approximation. We expand the basis so that indexes $\mu,\nu$
run over the spin ($s=\uparrow,\downarrow$) and valley ($\xi=\pm$) degrees of
freedom, giving a full interacting basis of dimension $d=8N_L$:
\begin{gather}
\label{eq:four-fermion Coulomb}
    \hat{\mathcal{H}}_{HF} = \sum_{\substack{\mathbf{k},\mu,\nu}}
    \big[\mathbb{K}_{\mu\nu}(\mathbf{k})+\mathbb{H}_{\mu\nu}(\mathbf{k})
    +\mathbb{F}_{\mu\nu}(\mathbf{k})\big]
    \hat{c}^\dagger_{\mu}(\mathbf{k}) \hat{c}_{\nu}(\mathbf{k})
\end{gather}
\begin{gather}
    \mathbb{H}_{\mu\nu}(\mathbf{k})= \frac{\delta_{\mu\nu}}{A}\,
    \mathcal{V}_C\!\left(\mathbf{0}\right)\sum_{\mathbf{p},\lambda}
    n_{\lambda\lambda}(\mathbf{p}) \\
    \mathbb{F}_{\mu\nu}(\mathbf{k})=-\frac{1}{A}\sum_{\mathbf{p}}
    \mathcal{V}_C\!\left(\mathbf{p}-\mathbf{k}\right) n_{\mu\nu}(\mathbf{p})
\end{gather}
where $\hat{c}$ and $\hat{c}^\dagger$ denote fermionic annihilation and
creation operators and
$n_{\alpha\beta}(\mathbf{q})=\langle\hat{c}^{\dagger}_{\beta}(\mathbf{q})
\hat{c}_{\alpha}(\mathbf{q})\rangle$ is the self-consistent density matrix. We
employ a dual-gated Coulomb potential
$\mathcal{V}_{C}(\mathbf{q}) = \frac{2\pi e^2}{\epsilon |\mathbf{q}|}
\tanh\left(d_{g}|\mathbf{q}|\right)$, with $\epsilon=10$ the dielectric
constant and $d_{g}=200$~\AA{} the gate distance. The Hartree term involves only the $\mathbf{q}=\mathbf{0}$ component of the potential and is cancelled by the electrostatic contribution of the gates, amounting to a uniform shift absorbed in the chemical potential; the symmetry breaking discussed in this work is therefore driven entirely by the exchange channel. The total energy is evaluated as
\begin{equation}
E=\sum_{\mathbf{k}}\operatorname{tr}
\left[\left(\mathbb{K}+\tfrac{1}{2}\,\mathbb{F}\right)n(\mathbf{k})\right],
\end{equation}
The HF iteration converges to a local minimum of the energy functional in the
vicinity of its initial condition. 
We employ six deterministic ans\"atze, all constructed in the same way: the
non-interacting Hamiltonian is diagonalized in each
valley, and the doping density is distributed between the two valleys with a
fixed fraction $f$,
\begin{equation}
n_{K}=f\,n_e,
\qquad
n_{K'}=(1-f)\,n_e,
\qquad
f\in\left\{\tfrac{1}{2},\ \tfrac{9}{16},\ \tfrac{5}{8},\ \tfrac{3}{4},\
\tfrac{7}{8},\ 1\right\}.
\end{equation}
This ladder of fractions interpolates from the valley-symmetric metal
($f=1/2$), which also provides the reference state, to the fully
valley-polarized one ($f=1$), passing through the partially
imbalanced and three-quarter states. Because the fractions are fixed, the
basin coverage is identical at every point of the diagram, eliminating the
spatial noise inherent to random initializations. No spin structure is imposed
by the seeds: the spin order---collinear locking or Rashba texture---is
selected spontaneously by the self-consistency, which is unbiased since
$\lambda_4$ already breaks the spin-rotation symmetry explicitly.

At every iteration the bands are refilled to the exact target density; when
the Fermi level falls inside a degenerate multiplet of $g$ states, all of them
are occupied with equal fractional weight (ensemble average), so that every
ansatz converges to exactly the same filling and total energies are directly
comparable. The density matrix is updated with linear mixing
\begin{equation}
n^{(i+1)}_{\mathrm{in}}=\alpha\, n^{(i)}_{\mathrm{out}}
+(1-\alpha)\, n^{(i)}_{\mathrm{in}}
\end{equation}
which damps the oscillations between nearly degenerate SOC-split
configurations. $\alpha=0.7$ was used in our calculation. The iteration terminates by satisfying our convergence criteria: either when the density-matrix residual
satisfies $\max|n_{\mathrm{out}}-n_{\mathrm{in}}|<10^{-7}$, or when the
relative change of the total energy remains below $10^{-10}$ over three
consecutive iterations; non-converged runs are discarded from the analysis.
The ground state at each point is the lowest-energy converged solution,
characterized by a condensation energy per doped carrier
\begin{equation}
E_{\mathrm{dope}}=\frac{E-E_{\mathrm{sym}}}{N_{d}},
\end{equation}
with $E_{\mathrm{sym}}$ the energy of the valley-symmetric reference at the
same filling and $N_{d}$ the number of doped carriers; a negative
$E_{\mathrm{dope}}$ signals a symmetry-broken solution. 
Calculations were performed at zero temperature on an hexagonal mesh of
$N_k=19\,441$ points around each valley, for each pair of electronic density
$n_e$ and displacement field $D$. \\

\clearpage
\section{Ground states classification: band polarization}\label{sec:sm-bandpol}

Because of the $\lambda_4$ term, $S_z$ is no longer conserved and the isospin labels in the spin and valley subspaces no longer characterize the ground
state. The model does preserve the valley $U(1)_\xi$ symmetry, so the
Hartree--Fock Hamiltonian stays block-diagonal in the valley index; diagonalizing
each block $H^{\xi}_{\rm HF}(\mathbf{k})$ gives the valley-resolved bands
$E^{\xi}_{m\mathbf{k}}$. The filling of band $m$ in valley $\xi$ is:
\begin{equation}
n^{\xi}_{m}=\frac{1}{N_k}\sum_{\mathbf{k}}\Theta\!\left(\mu-E^{\xi}_{m\mathbf{k}}\right),
\end{equation}
which equals $1$ for a full band and $0$ for an empty one. A band is metallic
when it is partially filled, and the number of such bands in valley $\xi$ can be
counted as:
\begin{equation}
\Nb^{\xi}=\sum_{m}\Theta\!\left(n^{\xi}_{m}\right)\,\Theta\!\left(1-n^{\xi}_{m}\right)
\end{equation}
where the product of step functions equals $1$ only for partial filling,
$0<n^{\xi}_{m}<1$. Note that this counting treats electron and hole doping on an
equal footing: the carrier sign never enters the definition, but only selects
which bands deviate from integer filling. Upon electron doping the lowest
conduction bands acquire $n^{\xi}_{m}>0$, whereas upon hole doping the topmost
valence bands are partially emptied to $n^{\xi}_{m}<1$; in either case the doped
bands---and only these---satisfy $0<n^{\xi}_{m}<1$ and contribute to
$\Nb^{\xi}$, while full and empty bands drop out identically. The total band count is therefore:
\begin{equation}
\Nb=\Nb^{K}+\Nb^{K'}
\end{equation}
Being an integer, $\Nb$ is insensitive to any continuous cutoff and remains a
good quantum number in the presence of intrinsic SOC, which makes it the primary label of
the phase classification.\\
\begin{figure*}[h]
    \includegraphics[scale=0.32]{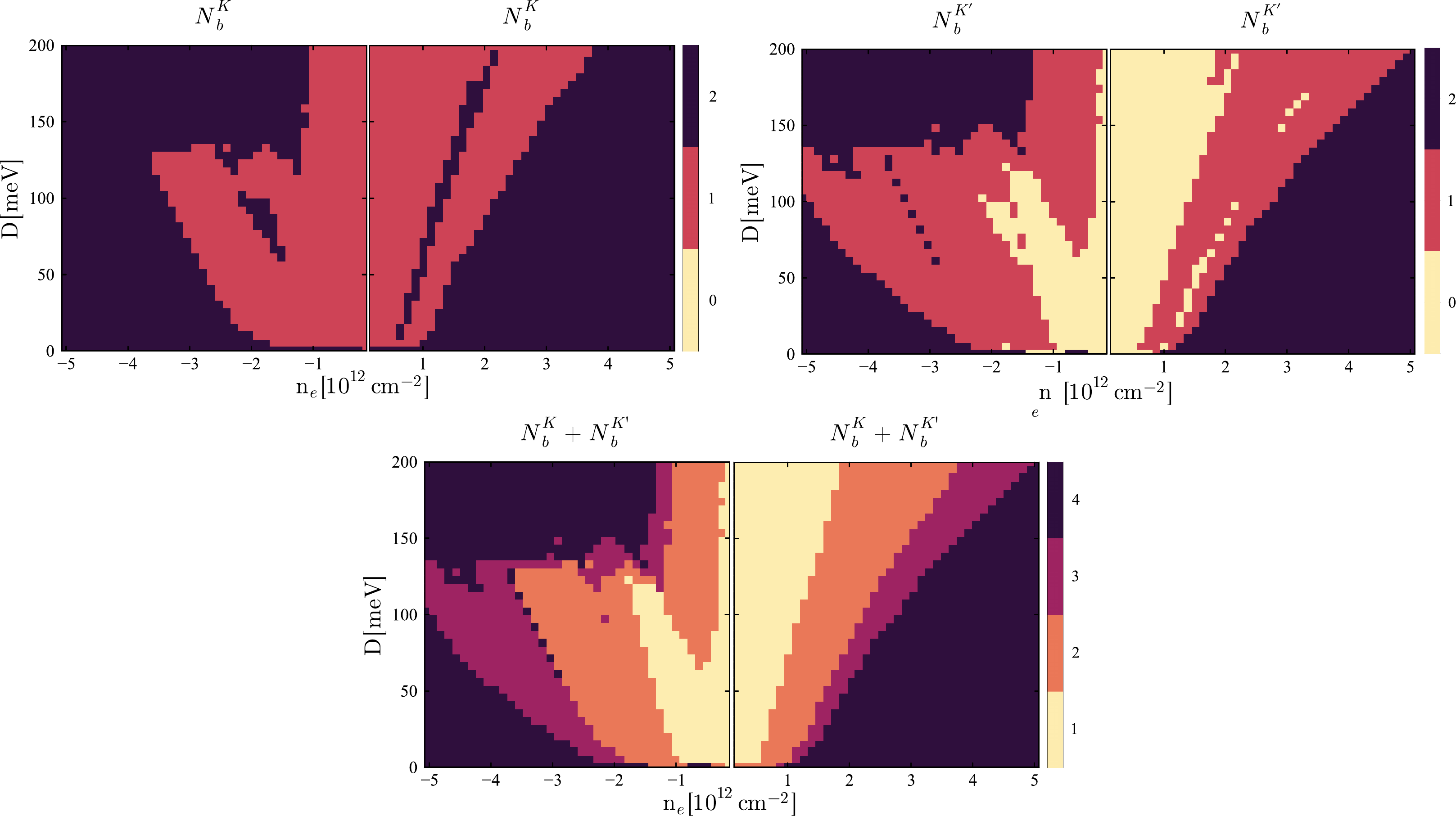}
    \caption{
    Band polarization phase diagram for Top) valley resolved $N_b^\xi$ for K valley (left) and K' valley (right) both for electron and hole doppings and  Bottom) total band polarization as the sum of both valleys.} 
    \label{fig:polband}
\end{figure*}

\clearpage
\section{Ground states classification: valley and spin--valley-locking polarization}\label{sec:sm-valleypol}

Since $\lambda_4$ breaks the spin-rotation symmetry explicitly, the spin
projection alone cannot label the interacting ground states. As introduced in
the main text, the classification is built on the band counting $\Nb^{\xi}$
supplemented by two flavor polarizations of the doped carriers, which we define
and map here. The carriers are described by the doped part of the
self-consistent density matrix, measured from charge neutrality,
\begin{equation}
\delta n_{\alpha\beta}(\mathbf{k})
= n_{\alpha\beta}(\mathbf{k})-n^{0}_{\alpha\beta}(\mathbf{k}),
\end{equation}
where $n^{0}$ is the density matrix of the charge-neutral system, so that
$\delta n$ contains only the added electrons or the removed holes. The number
of doped carriers is:
\begin{equation}
N_{d}=\sum_{\mathbf{k}}\operatorname{tr}\,\delta n(\mathbf{k}),
\end{equation}
negative for hole doping. The valley polarization is the population imbalance of the carriers between
the two valleys,
\begin{equation}
P_{v}=\frac{1}{N_{d}}\sum_{\mathbf{k}}
\operatorname{Tr}\!\left[\tau_{z}\,\delta n(\mathbf{k})\right],
\end{equation}
with $\tau_{z}$ the valley Pauli operator ($\tau_{z}=\pm1$ in valley
$\xi=\pm$). It ranges from $P_{v}=0$ for a balanced state to $|P_{v}|=1$ when
all carriers reside in a single valley. Since $\tau_{z}$ is odd under time
reversal, a finite $P_{v}$ signals the spontaneous breaking of $\mathcal{T}$:
these states carry a valley--orbital magnetization and are the ones accessible
to anomalous Hall or magneto-optical probes at zero field. The sign of
$P_{v}$ is selected spontaneously; all maps are shown after applying the
$\mathcal{T}$ gauge that brings the majority valley to $K$.\\

The spin--valley-locking polarization correlates the out-of-plane spin with
the valley,
\begin{equation}
P_{sv}=\frac{1}{N_{d}}\sum_{\mathbf{k}}
\operatorname{tr}\!\left[\tau_{z}\,s_{z}\,\delta n(\mathbf{k})\right],
\end{equation}
and reaches $|P_{sv}|=1$ when the carriers of valley $K$ are fully spin
polarized along $+\hat z$ and those of $K'$ along $-\hat z$ (or vice versa).
In contrast to $P_{v}$, the product $\tau_{z}s_{z}$ is even under
$\mathcal{T}$, so a large $P_{sv}$ is compatible with an unbroken time
reversal: it is the order parameter of the $\mathcal{T}$-symmetric H-SVL
phase. When a single valley is doped the two observables coincide with the
plain spin polarization, $P_{sv}\equiv P_{s}=\langle s_{z}\rangle$, which is
why the collinear quarter metal is labeled Q-$z$ rather than Q-SVL (see Table I from the main text).\\

\begin{figure*}[h]
    \includegraphics[scale=0.3]{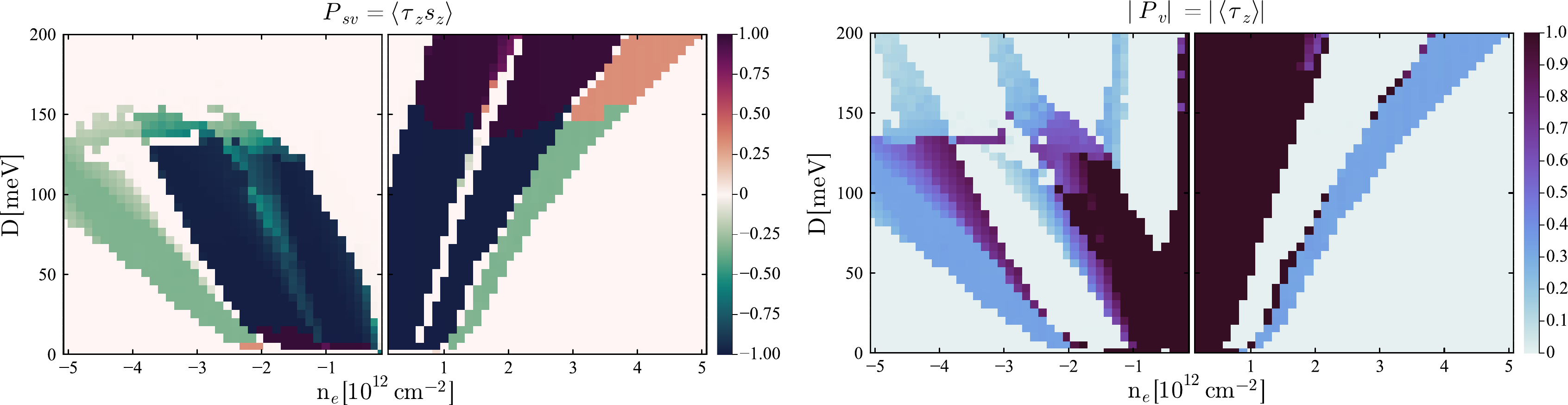}
    \caption{Valley polarization (left) and spin-valley-locking polarization (right) over the $(n_{e},D)$
plane for hole and electron doping. The $P_{v}$ maps organize the
$\mathcal{T}$-breaking phases: $P_{v}$ vanishes identically across F-R,
F-SVL, H-SVL and H-R, grows continuously through the valley-imbalanced (VI)
and three-quarter wedges, and saturates to $|P_{v}|=1$ in the H-V and Q
regions. The $P_{sv}$ maps are instead sharply bimodal: $|P_{sv}|\approx1$
in H-SVL and Q-$z$, $|P_{sv}|\approx1/3$ in TQ-SVL and $|P_{sv}|\approx0$ across all -R
phases, with no intermediate values in the balanced $(1,1)$ sector.}
\label{fig:polvalley}
\end{figure*}

\clearpage

\section{Fermi surface classification}\label{sec:sm-fs}
The splitting $\lambda_4$ in addition to the trigonal warping ($\gamma_3$) term of the RMG family confers on each valley a rich variety of Fermi-surface (FS) shapes. 
To classify them, we assign a carrier occupation to every metallic band $j$ of valley $\xi$ as:
\begin{equation}
\rho^{\xi}_{j}(\mathbf{k})=\Theta\!\left[h\,(\mu-E^{\xi}_{j\mathbf{k}})\right]
\end{equation}
with $h=+1$ for electron doping and $h=-1$ for hole doping, so that
$\rho^{\xi}_{j}=1$ on the filled conduction states or on the empty valence
states, respectively. The Fermi pocket of the band is the region
$\rho^{\xi}_{j}=1$, bounded by the FS contour $E^{\xi}_{j\mathbf{k}}=\mu$.

The topology of the pocket is captured by its Euler characteristic, computed on
the triangular $\mathbf{k}$ mesh from the occupations alone,
\begin{equation}
\chi^{\xi}_{j}
=\sum_{\mathbf{k}}\rho^{\xi}_{j}(\mathbf{k})
-\sum_{\langle\mathbf{k},\mathbf{k}'\rangle}
 \rho^{\xi}_{j}(\mathbf{k})\,\rho^{\xi}_{j}(\mathbf{k}')
+\sum_{\langle\mathbf{k},\mathbf{k}',\mathbf{k}''\rangle}
 \rho^{\xi}_{j}(\mathbf{k})\,\rho^{\xi}_{j}(\mathbf{k}')\,\rho^{\xi}_{j}(\mathbf{k}''),
\end{equation}
where the first sum runs over the mesh points, the second over nearest-neighbour
pairs, and the third over the elementary triangles of the mesh: the number of
vertices, minus bonds, plus triangles covered by the pocket. For any
two-dimensional region this integer equals the number of disconnected pockets
minus the number of holes they enclose. It therefore distinguishes all the FS
types realized in our phase diagram: $\chi=1$ for a simple circular pocket,
$\chi=3$ for the three warping pockets, $\chi=0$ for an annular FS (or an empty
band), and $\chi=4$ for four disjoint pockets. We evaluate $\chi^{\xi}_{j}$ for
each SOC-split band separately, so that pockets of one band lying inside the
other in projection are not lost; the construction makes no assumption of
$C_3$ symmetry. Since $\chi=0$ is ambiguous---annular versus empty---we combine it with the band
activity into a single fermiology index per band and valley,
\begin{equation}
\Lambda=
\begin{cases}
\,0  & \text{empty band}\\
\,1  & \chi=1 \quad(\text{circular})\\
\,3  & \chi=3 \quad(\text{three pockets})\\
\,7  & \chi=0 \quad(\text{annular})\\
\,12 & \chi=4 \quad(\text{four pockets}).
\end{cases}
\end{equation}
Each valley has two bands, a shallow band (the outer, first-to-empty
band, active for $\Nb^{\xi}\ge1$) and a deep band (its $\lambda_4$ partner,
active for $\Nb^{\xi}\ge2$). Thus, the total topological description for a valley is the sum of both:
\begin{equation}
\Lambda^{\rm tot}_{\xi}=\Lambda^{\rm sh}_{\xi}+\Lambda^{\rm dp}_{\xi}.
\end{equation}
The values $\{0,1,3,7,12\}$ are chosen so that the sum cannot equal degenerate values: the ten pairwise sums
$\{0,1,2,3,4,6,7,8,10,14\}$ are all distinct, so $\Lambda^{\rm tot}$ uniquely
decodes the pair of band topologies (Table~\ref{tab:lambda}). 
For the maps we use the compact rank
$g\in\{0,\dots,10\}$, the ascending rank of the realized $\Lambda^{\rm tot}$
values.

\begin{table}[h]
\centering\small
\setlength{\tabcolsep}{9pt} 
\begin{tabular}{ccccc}
\toprule
$g_\xi$ & $\Lambda^{\rm tot}_\xi$ & $(\Lambda^{\rm sh}_\xi,\Lambda^{\rm dp}_\xi)$ & $\Nb^{\xi}$ & Fermi Surface \\
\midrule
0  & 0  & $\left(0,0\right)$  & 0 & empty \\
1  & 1  & $\left(1,0\right)$  & 1 & simple circular  \\
2  & 2  & $\left(1,1\right)$  & 2 & double circular  \\
3  & 3  & $\left(3,0\right)$  & 1 & simple 3-pockets\\
4  & 4  & $\left(1,3\right)$  & 2 & circular $+$ 3-pockets \\
5  & 6  & $\left(3,3\right)$  & 2 & double 3-pockets  \\
6  & 7  & $\left(7,0\right)$  & 1 & simple annular  \\
7  & 8  & $\left(7,1\right)$  & 2 & annular $+$ circular \\
8  & 10 & $\left(7,3\right)$  & 2 & annular $+$ pockets \\
9  & 12 & $\left(12,0\right)$ & 1 & simple 4-pockets  \\
10 & 14 & $\left(7,7\right)$  & 2 & double annular\\
\bottomrule
\end{tabular}
\caption{Fermi surface per valley classification.}
\label{tab:lambda}
\end{table}

\begin{figure*}[h]
    \includegraphics[scale=0.3]{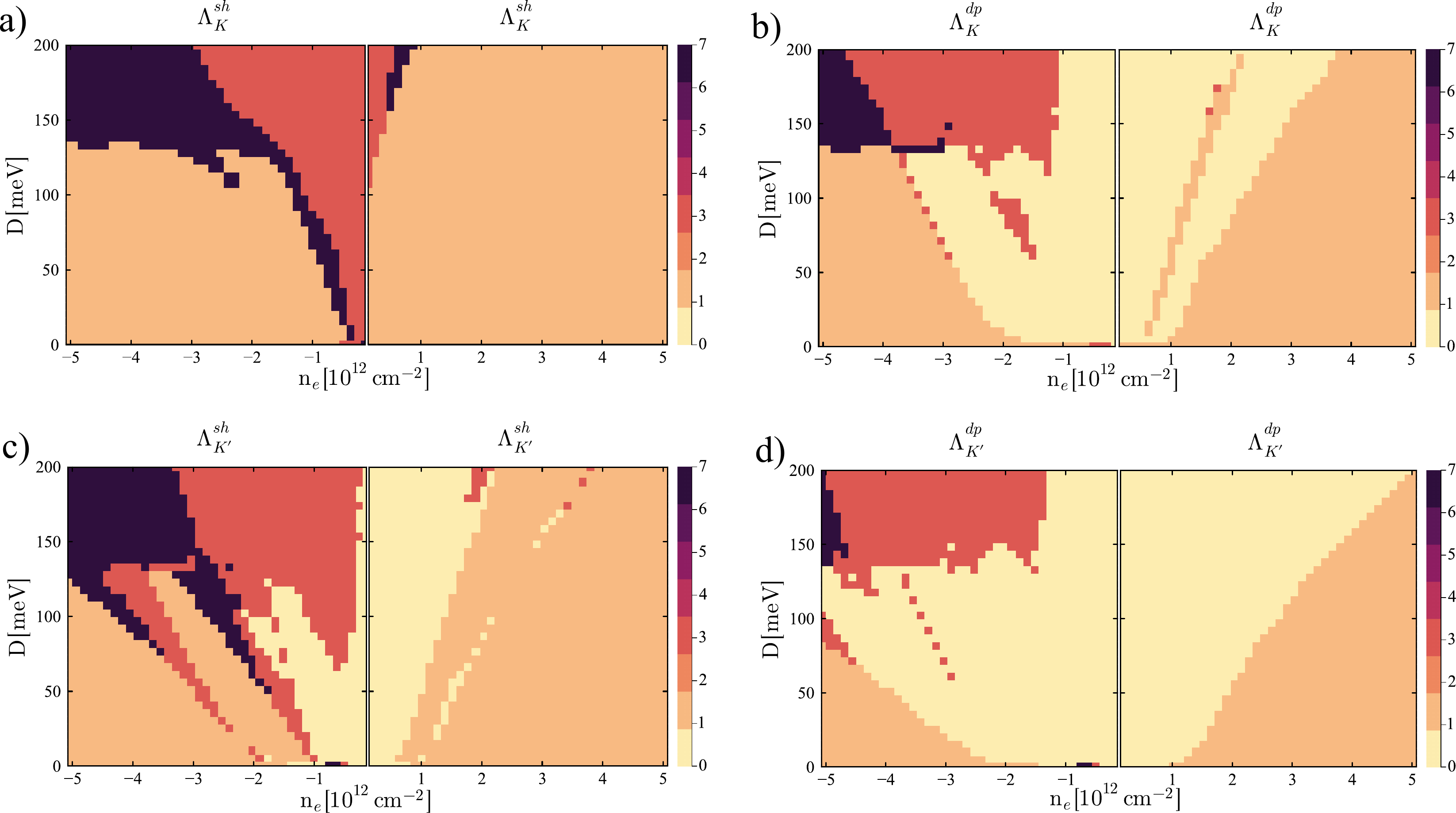}
    \caption{Renormalized Euler characteristic $\Lambda_\xi^j$ for a) K valley shallow band, b) K valley deep band, c) K' valley shallow band, d) K' valley deep band.}
\label{fig:euler}
\end{figure*}

\clearpage

\section{Response to a tilted magnetic field.}
\label{sec:tilted}
We analyze the hamiltonian which describes the response of flat bands of a rhombohedral stack to a tilted applied field. The hamiltonian can be found in eq.(\ref{eq:hamil_magf}). The $2 \times 2$ which describes the hamiltonian is:, for a given valley and sublattice
\begin{align}
    {\cal H}_{SOC} ( \vec{k} ) &= \lambda_1 \left( \begin{array}{rl} 1 &0 \\0 &- 1\end{array} \right) + \frac{\lambda_4 v_F}{t_\perp}\left( \begin{array}{lr} 0 &k_x + i k_y \\ k_x - i k_y &0 \end{array} \right) +  \mu_B \left( \begin{array}{lr} B_z &B_x \\ B_x &-B_z \end{array} \right)
    \label{eq:hamil_tilted}
\end{align}
First order perturbation theory on $B_x$ gives:
\begin{align}
    \langle s_x \rangle &= \mu_B B_x \sum_{\vec{k}} \left( \frac{1}{\lambda_1} + \frac{ t_\perp}{v_F \lambda_4} \frac{1}{| \vec{k} |} \right)
\end{align}
where the sum over $\vec{k}$ is taken over  occupied states, $| \vec{k} | \le k_F$, where $k_F$ is the Fermi wavevector. 
\\ \indent
A full calculation, assumming $B_z = B_0$ and $B_x = B_0 \tan ( \theta )$ gives:
\begin{align}
    \langle s_z \rangle &= \int_0^{k_F} k dk d \phi  \frac{\lambda_1 + \mu_B B_0}{\sqrt{(\lambda_1 + \mu_B B_0 )^2 + \left| \mu_B B_0 \tan ( \theta ) + \frac{\lambda_4 v_F k e^{i \phi}}{t_\perp} \right|^2}} \nonumber \\
    \langle s_x \rangle &= \int_0^{k_F} k dk d \phi  \frac{\mu_B B_0 \tan ( \theta ) + \frac{\lambda_4 v_F k \cos ( \phi )}{t_\perp}}{\sqrt{(\lambda_1 + \mu_B B_0 )^2 + \left| \mu_B B_0 \tan ( \theta ) + \frac{\lambda_4 v_F k e^{i \phi}}{t_\perp} \right|^2}}
     \label{spin_SSH_1}
\end{align}
For $( \lambda_4 v_F k ) / t_\perp \gg \lambda_1, \mu_B  B_0$ these equations reduce to:
\begin{align}
    \langle s_z \rangle &\approx \int_0^{k_F} k dk d \phi  \frac{\lambda_1 + \mu_B B_0}{\frac{\lambda_4 v_F k}{t_\perp}} \nonumber \\
    \langle s_x \rangle &\approx \int_0^{k_F} k dk d \phi  \frac{ \mu_B B_0 \tan ( \theta ) + \frac{\lambda_4 v_F k \cos ( \phi )}{t_\perp}}{\sqrt{\left( \frac{\lambda_4 v_F k}{t_\perp} \right)^2 + 2 \mu_B B_0 \tan ( \theta ) \frac{\lambda_4 v_F k \cos ( \phi )}{t_\perp}}} \approx \nonumber \\
    &\approx 
    \int_0^{k_F} k dk d \phi  \frac{\mu_B B_0 \tan ( \theta )[ 1 - \cos ( \phi )^2 ]}{ \frac{\lambda_4 v_F k}{t_\perp}}
\end{align}

In the limit $\lambda_4 \gg \lambda_1 , \mu_B | \vec{B} |$ The dependence of $\theta_S$ on $\theta_B$ can be approximated by the function:
\begin{align}
    \theta_S &= \arctan \left( \frac{\langle s_x \rangle}{\langle s_z \rangle} \right) =\arctan \left[ \frac{\mu_B B_0 \tan ( \theta_B )}{2 ( \lambda_1 + \mu_B B_0 )}\right]
    \label{spin_SSH_2}
   \end{align}

\begin{figure}
   {\includegraphics[scale=0.60]{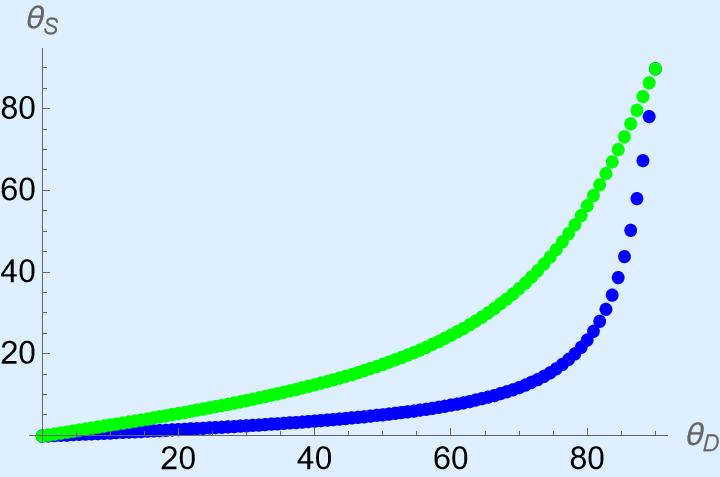}} 
   \caption{As in Fig.~\ref{fig:tilted} of the main text, but using the simplified model described by equations (\ref{spin_SSH_1}) and (\ref{spin_SSH_2}). The values of $\lambda_{at}$ used are $\lambda_{at} = 20$ meV, and $\lambda_{at} = 50$ meV, leading to $\lambda_1 = 0.013$ meV (green), and $\lambda_1 =0.081$ meV (blue).}
  \label{fig:comp_SSH}
\end{figure}

This limit is already reached for the parameters which approximate the experiments in Ref.~\cite{auerbach_isospin_2025}, see Fig.~[\ref{fig:comp_SSH}].

\clearpage  

\section{Simplified model. }
\label{sec:simplified}

A widely used simplified model for the electronic structure in the absence of spin-orbit coupling is defined in terms of only two parameters, the Fermi velocity, $v_F$, and the leading interlayer hopping, $\gamma_1 \approx 0.4$ eV. The bulk hamiltonian for a given valley is:
\begin{align}
    {\cal H}_{\vec{k}_\parallel , k_z} &= \left( \begin{array}{cc} 0 & v_F k_+ + \gamma_1 e^{i k_z} \\ v_F k_- + \gamma_1 e^{-i k_z} \end{array}
    \right)
    \label{eq:hamil_SSH}
    \end{align}
\begin{figure}[h!]
\begin{center}
\begin{tabular}{c}
{\includegraphics[scale=1]{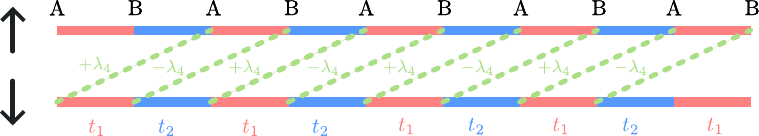}} 
\end{tabular}
\caption{Sketch of the SSH limit of rhombohedral graphene including the $\lambda_4$ term of the spin orbit coupling, see eq.~(\ref{eq:hamil_magf}) of the main text. The other parameters are $t_1 = v_F k$, and $t_2 = \gamma_1$.
}
\label{fig:sketch_SSH}
\end{center}
\end{figure}

This $\vec{k}_\parallel$ dependent one dimensional hamiltonian, is a version of the Su-Schrieffer-Heeger hamiltonian, with phases in the hopping terms. Adding the term in the spin-orbit coupling proportional to $\lambda_4$, we obtain the $4 \times 4$ hamiltonian, sketched in Fig.[\ref{fig:sketch_SSH}]:
\begin{align}
    \mathcal{H}_{\vec{k}_\parallel, k_z} &=
    \begin{pmatrix}
    0 & v_F k_+ + \gamma_1 e^{i k_z} & \lambda_4 e^{i k_z} & 0 \\[2pt]
    v_F k_- + \gamma_1 e^{-i k_z} & 0 & 0 & -\lambda_4 e^{i k_z} \\[2pt]
    \lambda_4 e^{-i k_z} & 0 & 0 & v_F k_+ + \gamma_1 e^{i k_z} \\[2pt]
    0 & -\lambda_4 e^{-i k_z} & v_F k_- + \gamma_1 e^{-i k_z} & 0
    \end{pmatrix}.
    \label{hamilSSH_v2}
\end{align}

The four bands are doubly degenerate:

\begin{align}
    \epsilon_{\vec{k}_{\parallel} , k_z} &=
    \pm \sqrt{\lambda_4^2 + \gamma_1^2 + v_F^2 | \vec{k}_{\parallel} |^2
    + 2 \gamma_1 v_F | \vec{k}_{\parallel} | \cos ( \phi_k - k_z )},
    \label{hamilSSH_spectra}
\end{align}

where $\phi_k$ defines the direction of $\vec{k}_\parallel$. 

\begin{figure}[h]
\begin{center}
\begin{tabular}{c}
{\includegraphics[scale=0.25]{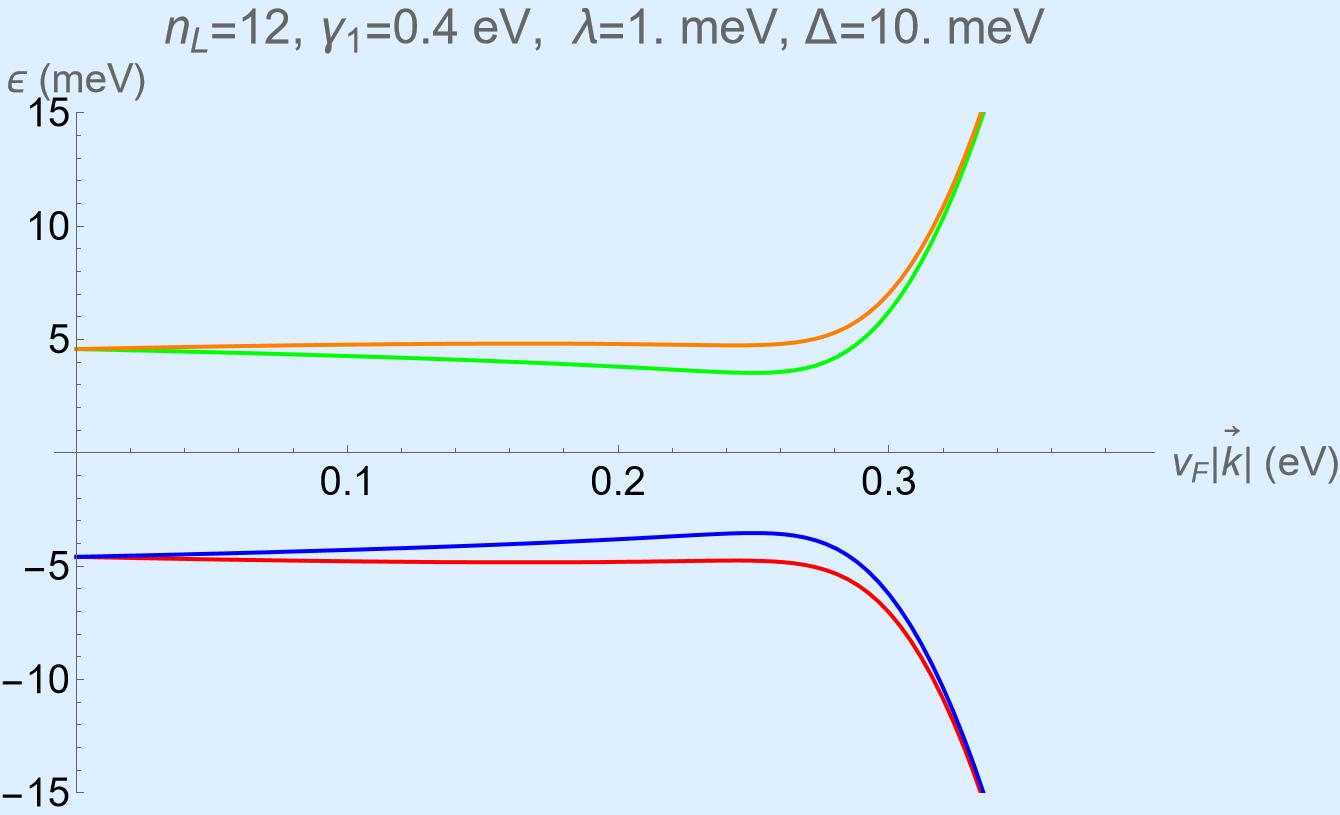}} \\
{\includegraphics[scale=0.25]{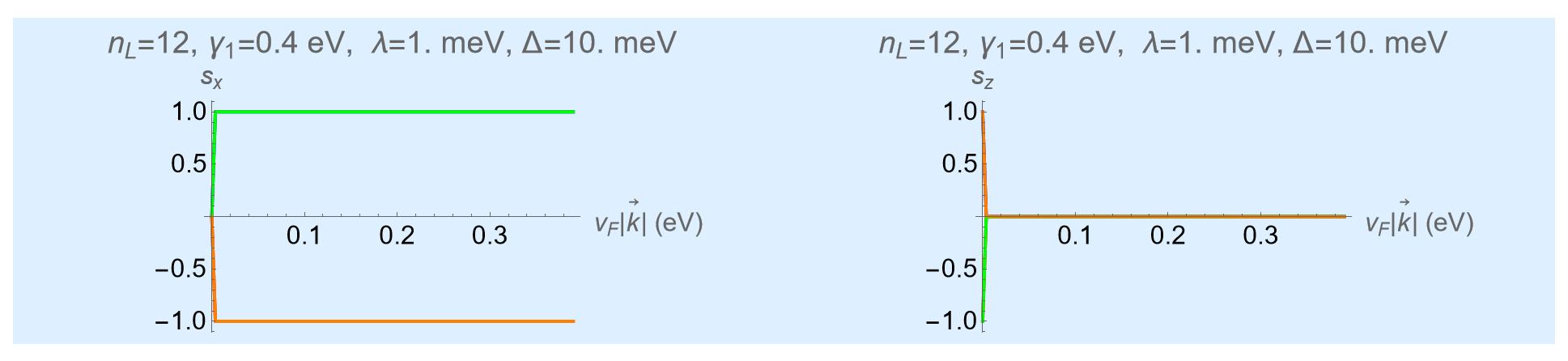}}
\end{tabular}
\caption{Top: central bands for a rhombohedral stack with $n_L = 12$ layers, spin-orbit coupling $\lambda_{4} = 1$ meV, and splitting between surfaces  $\Delta = 10$ meV. Bottom: values of the spins for the four central bands. The spins coincide in pairs of bands.
}
\label{fig:nl_12}
\end{center}
\end{figure}

\begin{figure}[h]
\begin{center}
\begin{tabular}{c}
{\includegraphics[scale=0.25]{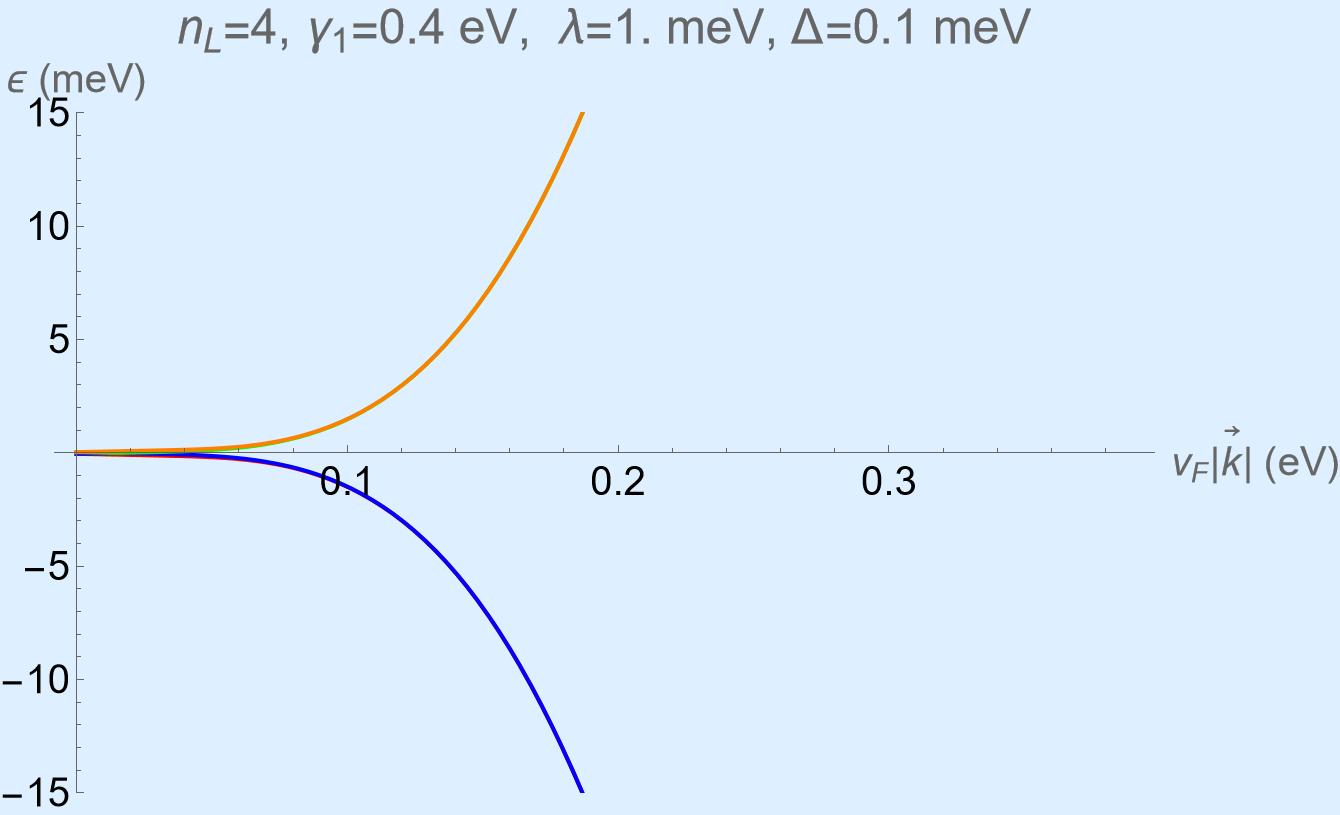}} \\
{\includegraphics[scale=0.20]{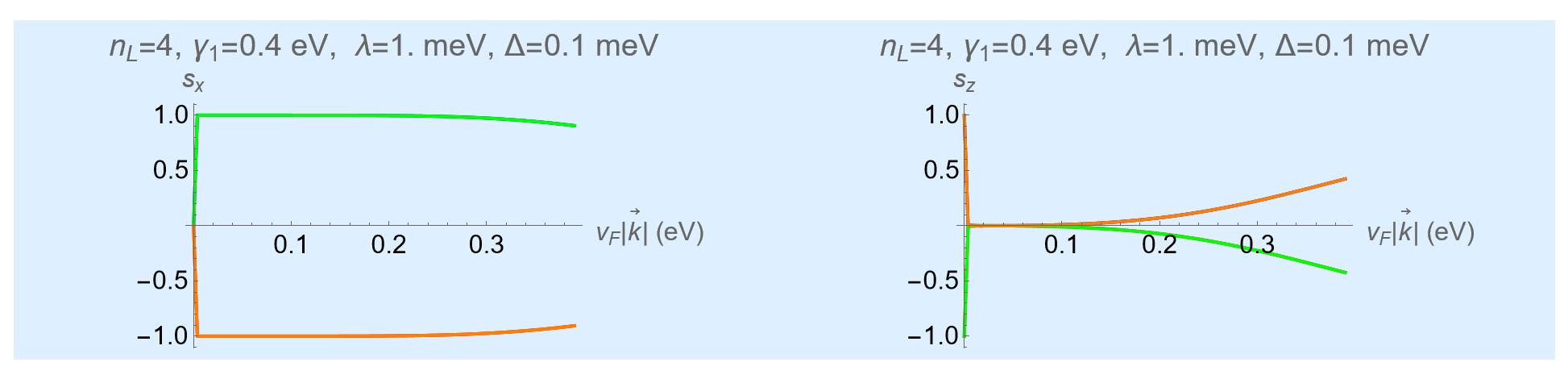}}
\end{tabular}
\caption{Top: central bands for a rhombohedral stack with $n_L = 4$ layers, spin-orbit coupling $\lambda_{4} = 1$ meV, and splitting between surfaces  $\Delta = 0.1$ meV. Bottom: values of the spins for the four central bands. The spins coincide in pairs of bands.
}
\label{fig:nl_4}
\end{center}
\end{figure}

\begin{figure}[h]
\begin{center}
\begin{tabular}{c}
{\includegraphics[scale=0.25]{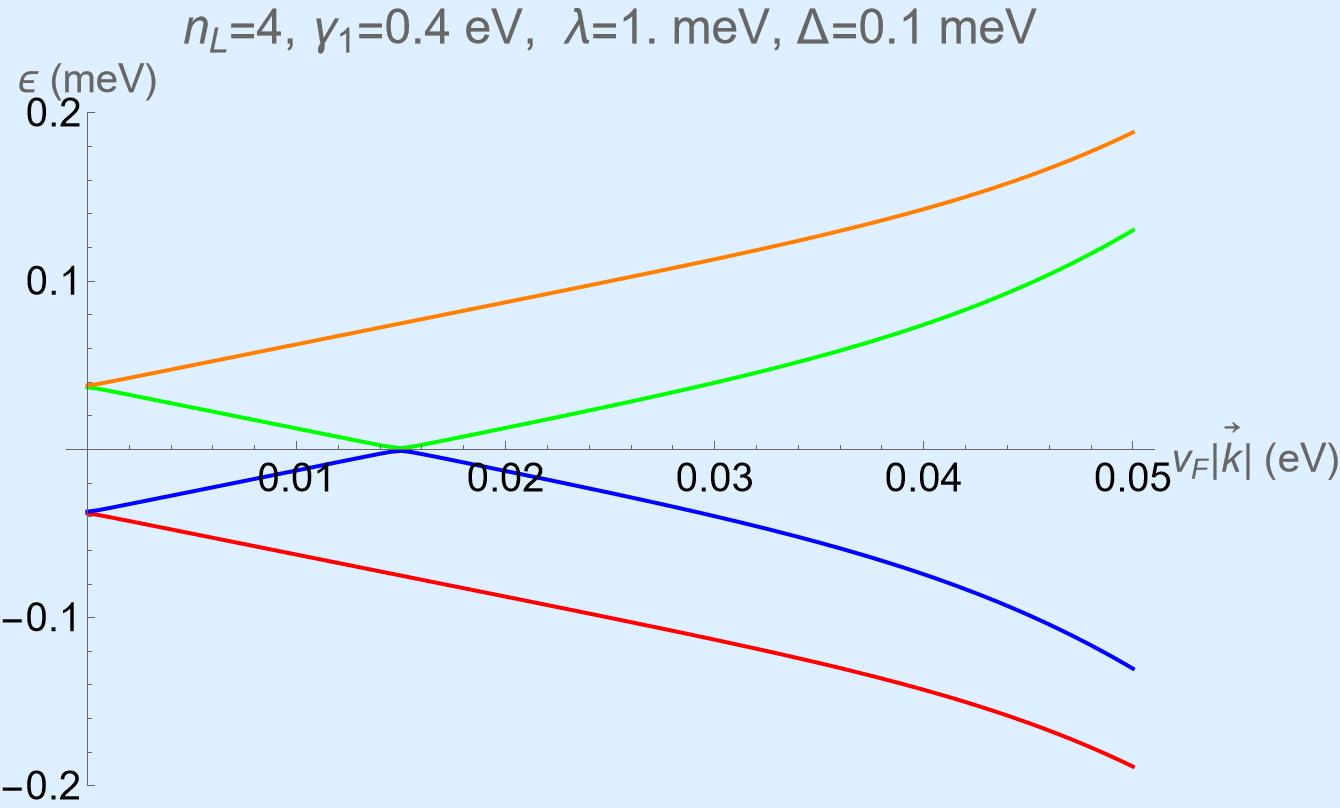}} 
\end{tabular}
\caption{Small $\vec{k}_\parallel$ behavior of the bands shown  in Fig.~[\ref{fig:nl_4}]}. 

\label{fig:nl_4_d}
\end{center}
\end{figure}

This model leads, in the absence of spin-orbit coupling and a displacement field, to four totally flat bands localized at the top and bottom surfaces, and centered at $K$ and $K'$, and with a radius $k = \gamma_1 / v_F$. The region in $\vec{k}_\parallel$ space where the flat band exists describes the situation where the SSH model has edge states. The combined effect of the spin-orbit coupling, $\lambda$,the displacement field, $\Delta$ leads, at least, to two different regimes:

i) For $N \gg 1$ and $\Delta \gg \lambda$ the four flat bands first split into two doublets, each localized at a given surface. These doublets are separated by en energy of order $\pm \Delta$. The spin-orbit coupling splits each of these doublets. The splitting depends on $\vec{k}_\parallel$, as the spin-orbit coupling hybridizes sites in different layers, and the amplitude of the wavefunction of the flat bands decays as $| \alpha_i | \propto [ ( v_F | \vec{k}_\parallel | ) / \gamma_1 ]^i$. A $2 \times 2$ effective hamiltonian can be defined for each of the surfaces of the stack:
\begin{align}
    {\cal H}_{eff} ( \vec{k}_\parallel ) &=  \frac{\lambda_4 v_F}{\gamma_1} \sigma_z\left( s_x k_y - s_y k_x \right)
    \label{hamildirac}
\end{align}
This is a Dirac hamiltonian where the spins lie within the $k_x - k_y$ plane and wind around the $K$ and $K'$ points. An in plane Zeeman field, $\delta {\cal H}_{x,y} = \lambda_Z B_{x,y}s_{x,y}$ plays the role of a constant gauge field, and an out of plane field, $\delta {\cal H}_{z} = \lambda_Z B_z s_{z}$, opens a gap at the Dirac point. An in plane field, $B_x$, displaces the Dirac point by $\Delta k \approx ( \gamma_1 \lambda_Z B_x ) / \lambda v_F )$ and it leads to a magnetization per atom in the flat bands of $m_x \sim ( \lambda_Z B_x ) /  \lambda $. An example of this regime is shown in Fig.~[\ref{fig:nl_12}]. 
Note that, including the Kane-Mele term previously discussed, proportional to $\lambda_1$, and an out of plane magnetic field, we obtain the hamiltonian:
\begin{align}
    {\cal H}_{eff}^{SOC} ( \vec{k}_\parallel ) &= \lambda_1 \tau_z \sigma_z s_z +  \frac{\lambda_4 v_F}{\gamma_1} \sigma_z\left( s_x k_y - s_y k_x \right)   + \mu_B B s_z
\label{eq:hamildirac_lambda_1}
\end{align}
For $\mu_B B > \lambda_1$ this hamiltonian describes a Quantum Hall insulator at neutrality, although with a small gap.

ii) For $\Delta \sim \lambda_4$ and $N$ not much larger that 1 the four flat bands need to be considered. The states in both surfaces become hybridized. In the absence of a displacement field the degeneracy of the four bands is not completely broken. Near $| \vec{k}_\parallel | = 0$ the bands can be described as two degenerate Dirac equations. The hybridization between states at the two surfaces is more pronounced at larger values of $| \vec{k}_\parallel |$, see Fig.~[\ref{fig:nl_4}]. At the edges of the flat bands the spin polarization acquires an out of plane component. Interestingly, the two sets of Dirac bands mentioned in i) cross, see Fig.~[\ref{fig:nl_4_d}].

\subsection{Analytic solutions for specific systems}

The simplified SSH-limit model introduced in the beginning of this section for rhombohedral stacks has analytic solutions for some graphitic systems. 
In this section, we provide the explicit construction of the spinful Hamiltonians for Bernal bilayer graphene (BBG) and rhombohedral trilayer graphene (RTG), including intralayer hopping, purely vertical interlayer coupling, $\gamma_{1}$, and spin–orbit interactions, $\lambda_{4}$. 
We present the Hamiltonians in a layer–sublattice–spin basis, describe the block structure that separates spin-conserving and spin-mixing terms, and give analytical expressions for the energy spectra. 
These expressions allow for a full characterization of the eigenvalues and the effect of spin–orbit coupling on the band structure.

The resulting bands, obtained from the analytical spectra of Eqs.~\eqref{eq:BBG_spectrum} and~\eqref{eq:RTG_spectrum} and shown in Figs.~\ref{fig:bands_SSH_BBG} and~\ref{fig:bands_SSH_RTG}, illustrate the same physical mechanism as the rhombohedral stack of Fig.~\ref{fig:nl_4}. The interlayer spin-orbit term $\lambda_{4}$ mixes opposite spins across adjacent layers and generates a linear-in-$k$ contribution to the low-energy dispersion with slope $\propto \lambda_{4}/\gamma_1$. For realistic values of $\lambda_{4} \sim 1$ meV this slope is small compared to $v_F|k|$, so the linear dispersion only becomes clearly visible when $\lambda_{4}$ approaches $\gamma_1$.

\subsection{Bernal bilayer graphene}

We first consider Bernal-stacked bilayer graphene (BBG) including a spin-dependent coupling. 
In the spinful sublattice-layer basis
\begin{equation}
\Psi =
\left(
A_{1\uparrow}, B_{1\uparrow}, A_{2\uparrow}, B_{2\uparrow},
A_{1\downarrow}, B_{1\downarrow}, A_{2\downarrow}, B_{2\downarrow}
\right)^T ,
\end{equation}
where $A,B$ denote the sublattices and $1,2$ label the layers, the Hamiltonian can be written in block form
\begin{equation}
\mathcal{H}_{\text{BBG}} =
\begin{pmatrix}
\mathcal{H}_{\text{BBG},\uparrow} & R_{\uparrow\downarrow} \\
R_{\uparrow\downarrow}^{\dagger} & \mathcal{H}_{\text{BBG},\downarrow}
\end{pmatrix}.
\end{equation}

The diagonal blocks describe the conventional spin-conserving Bernal bilayer Hamiltonian, while the off-diagonal block $R_{\uparrow\downarrow}$ contains the spin-orbit coupling that mixes opposite spins.

Explicitly, including the nearest-layer dimer hopping $\gamma_1$ and a spin-orbit coupling amplitude $\lambda_{4}$, the Hamiltonian reads
\begin{equation}
\mathcal{H}_{\text{BBG}} =
\begin{pmatrix}
0 & -k & 0 & \gamma_1 & 0 & 0 & 0 & 0 \\
-k & 0 & 0 & 0 & 0 & 0 & 0 & 0 \\
0 & 0 & 0 & -k & \lambda_{4} & 0 & 0 & 0 \\
\gamma_1 & 0 & -k & 0 & 0 & -\lambda_{4} & 0 & 0 \\
0 & 0 & \lambda_{4} & 0 & 0 & -k & 0 & \gamma_1 \\
0 & 0 & 0 & -\lambda_{4} & -k & 0 & 0 & 0 \\
0 & 0 & 0 & 0 & 0 & 0 & 0 & -k \\
0 & 0 & 0 & 0 & \gamma_1 & 0 & -k & 0
\end{pmatrix}.
\end{equation}

The terms proportional to $k$ describe intralayer Dirac hopping between $A$ and $B$ sites, while $\gamma_1$ couples dimer sites between adjacent layers. The parameter $\lambda_{4}$ introduces a spin-orbit interaction that couples equal sublattices with opposite spin.

The characteristic polynomial of the Hamiltonian factorizes as

\begin{equation}
\det(H-E I)=
\left(
k^4 + k^2(\lambda_{4}^2-2E^2) + E^2(-\lambda_{4}^2-\gamma_1^2+E^2)
\right)^2,
\end{equation}
showing that all eigenvalues are twofold degenerate.

Defining
\begin{equation}
E_{\pm}(k)=
\frac{1}{\sqrt{2}}
\sqrt{
2k^2+\lambda_{4}^2+\gamma_1^2
\pm
\sqrt{(\lambda_{4}^2+\gamma_1^2)^2+4k^2\gamma_1^2}
},
\end{equation}
the spectrum consists of four bands
\begin{equation}
\label{eq:BBG_spectrum}
E=\pm E_{+}(k),\quad \pm E_{-}(k),
\end{equation}
each of which is doubly degenerate.

\begin{figure*}[h]
\begin{center}
{\includegraphics[width=\columnwidth]{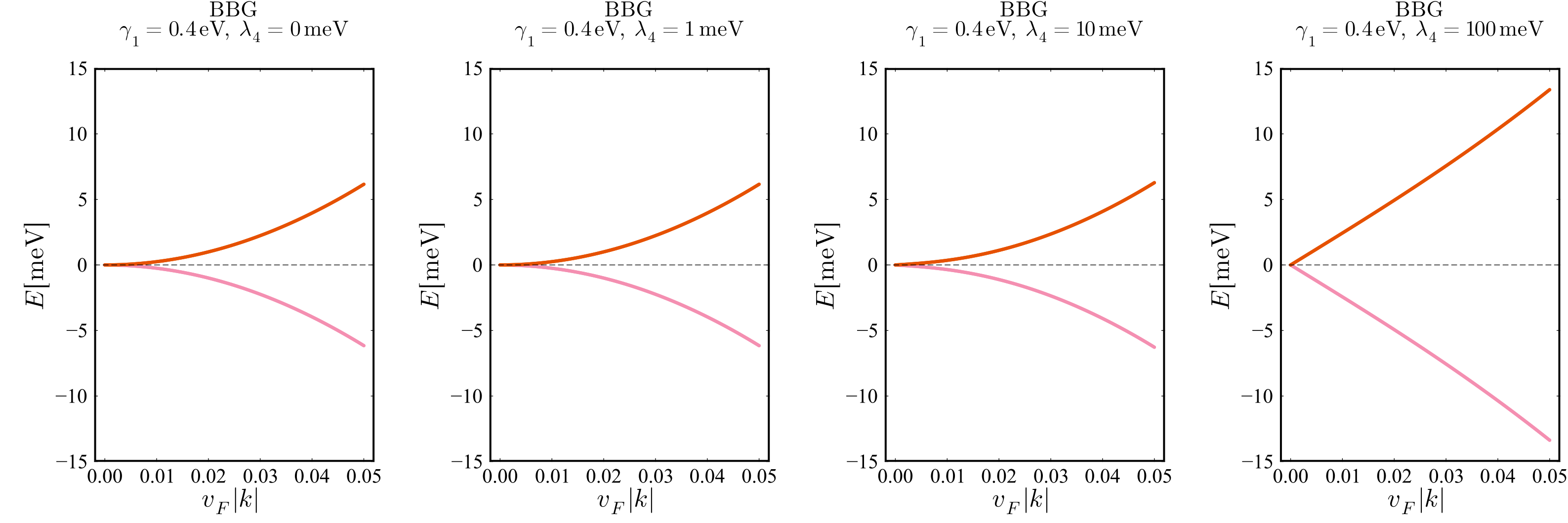}}
\caption{\textbf{The SSH limit in Bernal bilayer graphene.} Small-momentum band structure $E(k)$ obtained from the analytical spectrum of Eq.~\eqref{eq:BBG_spectrum}, for increasing values of the interlayer spin-orbit coupling $\lambda_{4} = 0, 1, 10, 100$ meV, with $\gamma_1 = 0.4$ eV and no displacement field. At $\lambda_{4} = 0$ the two low-energy bands touch quadratically at $E = 0$. The effect of $\lambda_{4}$ on the dispersion is very weak for realistic values of the coupling: up to $\lambda_{4} \sim 10$ meV the bands remain essentially indistinguishable from the $\lambda_{4} = 0$ case over this momentum range. Only for $\lambda_{4} \sim \gamma_1$ the quadratic touching gives way to a clearly linear crossing, whose slope grows monotonically with $\lambda_{4}$.}
\label{fig:bands_SSH_BBG}
\end{center}
\end{figure*}

The bands obtained from Eq.~\eqref{eq:BBG_spectrum} are shown in Fig.~\ref{fig:bands_SSH_BBG} for several values of $\lambda_{4}$. At $\lambda_{4} = 0$ the two low-energy bands touch quadratically at $E = 0$, with the usual gapless parabolic dispersion $E_{\pm}(k) \approx \pm k^{2}/\gamma_1$ characteristic of the Bernal bilayer. Turning on $\lambda_{4}$ breaks the twofold spin degeneracy and generates a linear-in-$k$ contribution. For realistic values of the interlayer spin-orbit coupling, $\lambda_{4} \sim 1$--$10$ meV $\ll \gamma_1$, the linear contribution is very small and the bands remain essentially parabolic and indistinguishable from the $\lambda_{4} = 0$ case over the plotted momentum range. The quadratic touching is only visibly converted into a linear crossing when $\lambda_{4}$ approaches $\gamma_1$, as can be seen in the rightmost panel of Fig.~\ref{fig:bands_SSH_BBG}. This shows that while $\lambda_{4}$ is the leading spin-orbit term in the multilayer, its effect on the band dispersion is quantitatively small unless the coupling is anomalously enhanced.

\subsection{Rhombohedral trilayer graphene}

We next consider rhombohedral trilayer graphene (RTG) with spin-orbit coupling. 
The spinful basis is
\begin{equation}
\Psi =
\left(
A_{1\uparrow}, B_{1\uparrow}, A_{2\uparrow}, B_{2\uparrow}, A_{3\uparrow}, B_{3\uparrow},
A_{1\downarrow}, B_{1\downarrow}, A_{2\downarrow}, B_{2\downarrow}, A_{3\downarrow}, B_{3\downarrow}
\right)^T .
\end{equation}

The Hamiltonian again takes the block form
\begin{equation}
\mathcal{H}_{\text{RTG}} =
\begin{pmatrix}
\mathcal{H}_{\text{RTG},\uparrow} & R_{\uparrow\downarrow} \\
R_{\uparrow\downarrow}^{\dagger} & \mathcal{H}_{\text{RTG},\downarrow}
\end{pmatrix}.
\end{equation}

The spin-conserving block is
\begin{equation}
\mathcal{H}_{\text{RTG},s}=
\begin{pmatrix}
0 & -k & 0 & \gamma_1 & 0 & 0 \\
-k & 0 & 0 & 0 & 0 & 0 \\
0 & 0 & 0 & -k & 0 & \gamma_1 \\
\gamma_1 & 0 & -k & 0 & 0 & 0 \\
0 & 0 & 0 & 0 & 0 & -k \\
0 & 0 & \gamma_1 & 0 & -k & 0
\end{pmatrix},
\end{equation}
while the spin-orbit coupling term reads
\begin{equation}
R_{\uparrow\downarrow}=
\begin{pmatrix}
0 & 0 & 0 & 0 & 0 & 0 \\
0 & 0 & 0 & 0 & 0 & 0 \\
\lambda_{4} & 0 & 0 & 0 & 0 & 0 \\
0 & -\lambda_{4} & 0 & 0 & 0 & 0 \\
0 & 0 & \lambda_{4} & 0 & 0 & 0 \\
0 & 0 & 0 & -\lambda_{4} & 0 & 0
\end{pmatrix}.
\end{equation}

The intralayer terms proportional to $k$ describe Dirac hopping within each graphene layer, while $\gamma_1$ couples dimer sites between adjacent layers in the rhombohedral stacking sequence. The parameter $\lambda_{4}$ introduces spin-orbit coupling between equal sublattices of opposite spin.

The eigenvalues of the Hamiltonian can be obtained analytically and may be written in terms of auxiliary quantities $A$, $B$, $C$, $C'$, and $D$ as

\begin{equation}
\label{eq:RTG_spectrum}
\begin{aligned}
E_{1,2} &= -\frac{\sqrt{C/B}}{\sqrt{6}}, \quad
E_{3,4} = \frac{\sqrt{C/B}}{\sqrt{6}}, \\
E_{5,6} &= -\frac{\sqrt{D/B}}{2\sqrt{3}}, \quad
E_{7,8} = \frac{\sqrt{D/B}}{2\sqrt{3}}, \\
E_{9,10} &= -\frac{\sqrt{C'/B}}{2\sqrt{3}}, \quad
E_{11,12} = \frac{\sqrt{C'/B}}{2\sqrt{3}},
\end{aligned}
\end{equation}
where the coefficients are defined as

\begin{subequations}
\begin{equation}
A = \sqrt{
4 \lambda_{4}^8 + \lambda_{4}^4 (13 k^2 + 16 \lambda_{4}^2) \gamma_1^2 
+ 2 (16 k^4 + 13 k^2 \lambda_{4}^2 + 12 \lambda_{4}^4) \gamma_1^4 
+ (13 k^2 + 16 \lambda_{4}^2) \gamma_1^6 + 4 \gamma_1^8
},
\end{equation}
\begin{equation}
B = \Bigl(
3 \, \mathrm{i} \sqrt{3} A k \gamma_1 
- (\lambda_{4}^2 + \gamma_1^2)\bigl(-9 k^2 \gamma_1^2 + 2 (\lambda_{4}^2 + \gamma_1^2)^2 \bigr)
\Bigr)^{1/3},
\end{equation}
\begin{equation}
\begin{split}
C &= 2 (2^{1/3}) \lambda_{4}^4 + 4 B \gamma_1^2 + 2 (2^{1/3}) \gamma_1^4 
+ 4 \lambda_{4}^2 (B + 2^{1/3} \gamma_1^2)
+ 6 k^2 (B + 2 (2^{1/3}) \gamma_1^2) \\
&\quad + \Bigl(
3 \, \mathrm{i} \sqrt{3} A k \gamma_1 
- (\lambda_{4}^2 + \gamma_1^2)(-9 k^2 \gamma_1^2 + 2 (\lambda_{4}^2 + \gamma_1^2)^2)
\Bigr)^{2/3},
\end{split}
\end{equation}
\begin{equation}
\begin{split}
C' &= -4 (-2)^{1/3} \lambda_{4}^4 + 8 B \gamma_1^2 - 4 (-2)^{1/3} \gamma_1^4 
+ 12 k^2 (B - 2 (-2)^{1/3} \gamma_1^2) 
+ 8 \lambda_{4}^2 (B - (-2)^{1/3} \gamma_1^2) \\
&\quad + \Bigl(
3 \, \mathrm{i} \sqrt{3} A k \gamma_1 
- (\lambda_{4}^2 + \gamma_1^2)(-9 k^2 \gamma_1^2 + 2 (\lambda_{4}^2 + \gamma_1^2)^2)
\Bigr)^{2/3},
\end{split}
\end{equation}
\begin{equation}
\begin{split}
D &= 4 (-1)^{2/3} 2^{1/3} \lambda_{4}^4 + 8 B \gamma_1^2 + 4 (-1)^{2/3} 2^{1/3} \gamma_1^4 
+ 8 \lambda_{4}^2 (B + (-1)^{2/3} 2^{1/3} \gamma_1^2) \\
&\quad + 12 k^2 (B + 2 (-1)^{2/3} 2^{1/3} \gamma_1^2) 
- \mathrm{i} (-\mathrm{i} + \sqrt{3}) 
\Bigl(
3 \, \mathrm{i} \sqrt{3} A k \gamma_1 
- (\lambda_{4}^2 + \gamma_1^2)(-9 k^2 \gamma_1^2 + 2 (\lambda_{4}^2 + \gamma_1^2)^2)
\Bigr)^{2/3}.
\end{split}
\end{equation}
\end{subequations}

These expressions provide the full set of twelve eigenvalues of the spinful RTG Hamiltonian.  

\begin{figure*}[h]
\begin{center}
{\includegraphics[width=\columnwidth]{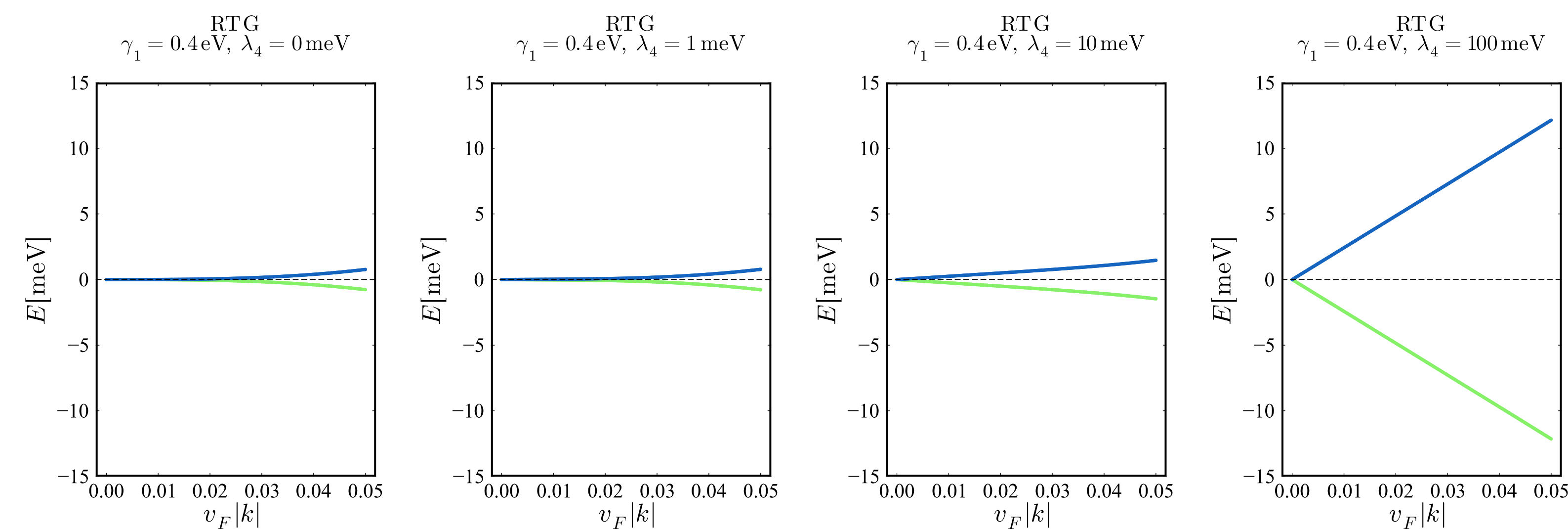}}
\caption{\textbf{The SSH limit in rhombohedral trilayer graphene.} Small-momentum band structure $E(k)$ obtained from the analytical spectrum of Eq.~\eqref{eq:RTG_spectrum}, for increasing values of the interlayer spin-orbit coupling $\lambda_{4} = 0, 1, 10, 100$ meV, with $\gamma_1 = 0.4$ eV and no displacement field. At $\lambda_{4} = 0$ the surface bands are nearly flat at $E = 0$, dispersing only cubically in $k$. As in BBG, the effect of $\lambda_{4}$ on the dispersion is very weak for realistic values of the coupling: up to $\lambda_{4} \sim 10$ meV the bands remain essentially flat over this momentum range. Only for $\lambda_{4} \sim \gamma_1$ the surface bands split into two clearly linear Dirac-like pairs with a slope that grows monotonically with $\lambda_{4}$.}
\label{fig:bands_SSH_RTG}
\end{center}
\end{figure*}

The bands obtained from Eq.~\eqref{eq:RTG_spectrum} are shown in Fig.~\ref{fig:bands_SSH_RTG} for several values of $\lambda_{4}$. At $\lambda_{4} = 0$ the low-energy bands are the surface states of the SSH chain, which disperse only cubically in $k$ and are therefore nearly flat over the plotted range. A finite $\lambda_{4}$ hybridizes the two surfaces across the stack and generates a linear-in-$k$ contribution, as in BBG. For realistic couplings, $\lambda_{4} \lesssim 10$ meV, this linear contribution is very small and the bands remain essentially flat and indistinguishable from the $\lambda_{4} = 0$ case. Only when $\lambda_{4}$ becomes comparable to $\gamma_1$ do the surface bands split into two clearly linear Dirac-like pairs, with a well-resolved spin splitting at finite $k$, as can be seen in the rightmost panel of Fig.~\ref{fig:bands_SSH_RTG}. The main qualitative difference with respect to BBG is that RTG already hosts two nearly degenerate surface bands at $\lambda_{4} = 0$, while BBG has a single quadratic touching that only converts into a linear crossing once $\lambda_{4} \neq 0$.

\end{document}